\documentclass[aps, floats, preprintnumbers, prl, groupedaddress, reprint,superscriptaddress,floatfix]{revtex4-1}

\usepackage{physics,slashed,amsmath,amssymb,epsfig,color,bbold,ulem,xcolor}
\allowdisplaybreaks

\newcommand{\g}{\eta}

\newcommand{\iu}{{\rm i}}
\newcommand{\e}{{\rm e}}

\newcommand{\nl}{\nonumber \\ }

\usepackage{feynmp-auto}
\newcommand{\marrow}[5]{%
    \fmfcmd{style_def marrow#1
    expr p = drawarrow subpath (1/4, 3/4) of p shifted 6 #2 withpen pencircle scaled 0.4;
    label.#3(btex #4 etex, point 0.5 of p shifted 6 #2);
    enddef;}
    \fmf{marrow#1,tension=0}{#5}}

\newcommand{\noarrow}[5]{%
    \fmfcmd{style_def marrow#1
    expr p = drawarrow subpath (1/2, 1/2) of p shifted 6 #2 withpen pencircle scaled 0.4;
    label.#3(btex #4 etex, point 0.5 of p shifted 6 #2);
    enddef;}
    \fmf{marrow#1,tension=0}{#5}}

\newcommand{\Em}{E_m}

\usepackage{graphicx}
\usepackage{epstopdf}
\usepackage{color}
\definecolor{orange}{RGB}{255,127,0}
\definecolor{blue2}{RGB}{33,114,173}
\begin{document}

\begin{fmffile}{feynmffile} 
\fmfcmd{%
vardef middir(expr p,ang) = dir(angle direction length(p)/2 of p + ang) enddef;
style_def arrow_left expr p = shrink(.7); cfill(arrow p shifted(4thick*middir(p,90))); endshrink enddef;
style_def arrow_left_more expr p = shrink(.7); cfill(arrow p shifted(6thick*middir(p,90))); endshrink enddef;
style_def arrow_right expr p = shrink(.7); cfill(arrow p shifted(4thick*middir(p,-90))); endshrink enddef;}

\fmfset{arrow_ang}{15}
\fmfset{arrow_len}{2.5mm}
\fmfset{decor_size}{3mm}

\preprint{FERMILAB-PUB-23-453-T}
\preprint{CALT-TH/2023-029}

\title{Field theory of the Fermi function}
\author{Richard J. Hill}
\email{richard.hill@uky.edu}
\affiliation{University of Kentucky, Department of Physics and Astronomy, Lexington, KY 40506, USA}
\affiliation{Fermilab, Theoretical Physics Department, Batavia, IL 60510, USA}

\author{Ryan Plestid}
\email{rplestid@caltech.edu}
\affiliation{University of Kentucky, Department of Physics and Astronomy, Lexington, KY 40506, USA}
\affiliation{Fermilab, Theoretical Physics Department, Batavia, IL 60510, USA}
\affiliation{Walter Burke Institute for Theoretical Physics, California Institute of Technology, Pasadena, CA, 91125 USA}

\begin{abstract}
  The Fermi function $F(Z,E)$ accounts for QED corrections to
  beta decays that are enhanced at either small electron velocity $\beta$ 
  or large nuclear charge $Z$.  For precision applications, the Fermi 
  function must be combined with other radiative corrections
  and with scale-  and scheme-dependent hadronic matrix elements.
  We formulate the Fermi function as a field theory object and
  present a new factorization formula for 
  QED radiative corrections to beta decays. 
  We provide new results for the anomalous dimension of the corresponding
  effective operator complete through three loops, 
  and resum perturbative logarithms and $\pi$-enhancements
  with renormalization group methods. 
  Our results are important for tests of fundamental physics
  with precision beta decay and related processes. 
\end{abstract}

\maketitle

\noindent{\bf Introduction.}
Many precision measurements and new physics searches involve charged
leptons interacting with nucleons or nuclei.  Examples include neutrino
scattering to obtain fundamental neutrino parameters~\cite{Nunokawa:2007qh,Hyper-KamiokandeWorkingGroup:2014czz,Diwan:2016gmz,NOvA:2021nfi,T2K:2021xwb,DUNE:2020jqi}; muon-to-electron conversion to search for
charged lepton flavor violation~\cite{deGouvea:2013zba,Bernstein:2013hba,Lee:2018wcx,Bernstein:2019fyh}; and beta decay to measure fundamental
constants~\cite{Bopp:1986rt,Ando:2004rk,Darius:2017arh,Seng:2018yzq,Seng:2018qru,Fry:2018kvq,Czarnecki:2019mwq,Hayen:2020cxh,Seng:2020wjq,Gorchtein:2021fce,Hardy:2020qwl,UCNt:2021pcg,Shiells:2020fqp} and search for new physics~\cite{Cirigliano:2013xha,Glick-Magid:2016rsv,Gonzalez-Alonso:2018omy,Glick-Magid:2021uwb,Falkowski:2021vdg,Brodeur:2023eul,Crivellin:2020ebi,Coutinho:2019aiy,Crivellin:2021njn,Crivellin:2020lzu,Cirigliano:2022yyo}.
It is important to control radiative
corrections to these processes~\cite{Sirlin:1967zza,Jaus:1972hua,Wilkinson:1982hu,Sirlin:1986cc,Jaus:1986te,Cirigliano:2022hob}.  
The precision demands of superallowed nuclear beta decays are particularly stringent.  As a consequence of Conserved Vector Current (CVC) relations, hadron and nuclear structure enter as small corrections.      
Experimental and nuclear uncertainties are being pushed to the level of 100 ppm~\cite{Hardy:2020qwl,Gorchtein:2023naa}, providing the best determination of the fundamental Cabibbo-Kobayashi-Maskawa (CKM) quark mixing parameter $|V_{ud}|$, and the most stringent low-energy constraint on scalar currents beyond the Standard Model~\cite{Hardy:2020qwl}. 
%
%
In this {\it Letter} we present new results for long-distance QED corrections to beta decay~\cite{z2a3anom,Hill:2023bfh,largepi}, and discuss implications for the CKM unitarity discrepancy~\cite{Hardy:2020qwl} 
and New Physics constraints.   

QED 
corrections are dramatically enhanced relative to naive power counting
in the fine structure constant $\alpha \approx 1/137$
for large-$Z$ nuclei and for small-$\beta$ leptons
($Z$ denotes the nuclear charge,
and $\beta$ the lepton velocity).  
The Fermi function~\cite{Fermi:1934hr} in beta decay 
describes the enhancement (suppression) for negatively (positively) charged leptons propagating in a nuclear Coulomb field.  For a nuclear charge $Z$ and electron energy $E$ it is traditionally defined by solving the Dirac equation in a point-like Coulomb field. The result is then given as~\cite{Fermi:1934hr,Hayen:2017pwg}:
\begin{align}\label{eq:fermi}
  F(Z,E) &= { 2(1+\g) \over | \Gamma(2\g + 1)|^2} [\Gamma(\g+ \iu \xi)]^2 \e^{\pi \xi} (2pr)^{2(\g-1)}
    \,,
\end{align}
where $\g \equiv \sqrt{1-(Z\alpha)^2}$, $\xi=Z\alpha/\beta$, $p=\sqrt{E^2-m^2}$
and  $m$ is the electron mass. 
The quantity $r$ denotes a short distance regulator identified approximately as the nuclear
size~\footnote{The ambiguity in defining $r$ is often resolved by including an explicit nuclear charge density distribution~\cite{Hayen:2017pwg}. This procedure does not address how to interface the Fermi function with EFT matching conditions, or how to properly interface with 
higher-order radiative corrections.}.
Several questions arise in the application of $F(Z,E)$ to
physical processes: 1) What is the scale $r^{-1}$ and how does it relate to
conventional renormalization in quantum field theory?
2) How can other radiative corrections be included systematically?
3) What is the relation between the Fermi function with $Z=1$ and
the radiative correction to neutron beta decay? 
Answering these questions is important for the interpretation of 
precision beta decay experiments. For example, 
corrections at order $\alpha(Z\alpha)^2$ 
must be included at the 
current precision ($\sim 3\times 10^{-4}$) 
of $|V_{ud}|$ extractions~\cite{Hardy:2020qwl}. 
These corrections 
require a theoretically self-consistent treatment of both the Fermi function and other radiative corrections, but have previously been treated only in a 
heuristic ansatz~\cite{Sirlin:1986cc,Sirlin:1986hpu}.
To answer these questions, we re-formulate the Fermi function in effective
field theory (EFT), and study its interplay with subleading radiative corrections. 



\noindent{\bf Factorization and all-orders matching.}
Factorization arises from the separation of different energy scales
involved in a physical process~\cite{Collins:1989gx,Bodwin:1994jh,Bauer:2002nz}. 
Nuclear beta decays involve physics at the weak scale $\sim 100~{\rm GeV}$, the hadronic scale $\sim 1~{\rm GeV}$, the scale of nuclear structure $\Lambda_{\rm nuc.} \sim 100~{\rm MeV}$, and the kinematic scales relevant for beta decay $E\sim 1~{\rm MeV}$.
The methods of EFT allow for each scale to be treated separately, and facilitate the calculation of higher order radiative corrections. 
In a sequence of EFTs, the components of a factorization formula are
identified with a corresponding sequence of matching coefficients, and a final low-energy matrix element. In the context of nuclear beta decays, the long-distance (or outer) radiative corrections can be computed in the low-energy effective theory, while structure dependent and short-distance (or inner) radiative corrections are absorbed into the Wilson coefficient. Real radiation is straightforwardly included~\footnote{See Supplemental Material [URL-HERE], which includes Ref.~\cite{Isgur:1990yhj}, for a discussion of the factorization theorem and how real radiation is included.}. 

Consider the  corrections to a
tree-level contact interaction with
a relativistic electron in the final state.
Ladder diagrams from a Coulomb potential with source charge $+Ze$ correct the tree level amplitude, ${\cal M}_{\rm tree}$, 
with explicit loop integrals given by (see Ref.~\cite{Hill:2023bfh} for more details)


%
\begin{widetext}
\begin{align}
\bar{u}(p) {\cal M} &= 
\label{eq:DCgeneral}
    \sum_{n=0}^\infty(Z e^2)^n 
    \int {\dd^D L_1 \over (2\pi)^D} \int {\dd^D L_2 \over (2\pi)^D}\cdots
    \int {\dd^D L_n \over (2\pi)^D}
        {1\over \vb{L}_1^2 + \lambda^2}{1\over (\vb{L}_1 -\vb{p})^2 -\vb{p}^2-\iu 0 }
          \nl &
         \hspace{0.075\linewidth}
         \times
        {1 \over (\vb{L}_1 - \vb{L}_2)^2 + \lambda^2} 
        {1\over (\vb{L}_2 -\vb{p})^2 -\vb{p}^2 -\iu 0} \cdots
        {1 \over (\vb{L}_{n-1} - \vb{L}_{n})^2 + \lambda^2} 
     {1\over (\vb{L}_n -\vb{p})^2 -\vb{p}^2 -\iu 0} 
    \\
    &\hspace{0.3\linewidth} \times
    \bar{u}(p) \gamma^0 ( \slashed{p} - \slashed{L}_1 + m) \gamma^0 (\slashed{p} - \slashed{L}_2 + m)  \cdots \gamma^0 ( \slashed{p} - \slashed{L}_n + m) {\cal M}_{\rm tree} \nonumber
    \,.
\end{align}
\end{widetext}
Integrals are evaluated in dimensional regularization with 
$D=3-2\epsilon$ dimensions, and we have included a photon mass, $\lambda$, to regulate infrared divergences~\footnote{
The parameter
$\lambda$ 
may be interpreted physically as 
the scale of atomic screening, which influences decay rates at sub-leading power~\cite{Hayen:2017pwg}.}.

In contrast to the non-relativistic problem \footnote{The classic Sommerfeld enhancement is obtained in a perturbative expansion by replacing the Dirac structures in the last line of (\ref{eq:DCgeneral}) with $2m$ in every propagator. This renders the problem UV finite \cite{Hill:2023bfh}.},  the
relativistic expression (\ref{eq:DCgeneral}) is UV divergent beginning at two-loop order,
indicating sensitivity to short-distance
structure.
The factorization theorem reads~\cite{Hill:2023bfh}
\begin{align}\label{eq:DCfac}
  {\cal M} = {\cal M}_{S}(\lambda/\mu_S) {\cal M}_H(p/\mu_S, p/\mu_H) {\cal M}_{\rm UV}(\Lambda/\mu_H) \,,
\end{align}
counting $p\sim m \sim E$ and where $\Lambda$ denotes the scale of hadronic and nuclear structure. We retain separate factorization scales $\mu_S$ and $\mu_H$ for clarity; conventional single scale matrix elements are obtained by setting $\mu_S=\mu_H=\mu$.
After ${\overline{\rm MS}}$ renormalization, to all orders in $Z\alpha$, the soft function is given by ${\cal M}_S = \exp\left( \iu \xi \log{\mu_S\over \lambda} \right)$~\cite{Yennie:1961ad,Weinberg:1965nx}. Our result for the hard function is new~\cite{Hill:2023bfh}, and is given (again to all orders in $Z\alpha$) by~%
\footnote{Covariant expressions are obtained by replacing $\beta\rightarrow \sqrt{1-E^2/m^2}$, 
$\gamma_0 \rightarrow v_\mu \gamma^\mu$, and $E\rightarrow v_\mu p^\mu$ where $v_\mu$ is the reference vector introduced in Eqs.\ (\ref{eq:superallowedL}) and (\ref{eq:neutronL}).}
\begin{widetext}
\begin{align}
    \label{eq:MHDren}
    \mathcal{M}_H
    &= e^{\tfrac{\pi}{2}\xi+\iu \phi_H}
            {2 \Gamma(\g-\iu \xi) \over \Gamma(2\g+1)}
    \sqrt{ \g - \iu \xi \over 1 - \iu \xi {m\over E} }  
   \sqrt{ E + \g m \over E + m} 
        {\sqrt{2 \g\over 1+\g}}     
   \left( 2p \over e^{\gamma_{\rm E}}\mu_H\right)^{\g-1}
   \bigg[ {1+\gamma^0\over 2} + {E+m\over E+\g m} \left( 1 - \iu \xi {m\over E}\right){1-\gamma^0\over 2} \bigg]
\,,
\end{align}
\end{widetext}
where $\phi_H=\xi\left(\log{2p\over\mu_S} - \gamma_{\rm E}\right) - (\g-1){\pi\over 2}$, $\gamma^0$ is a Dirac matrix, and $\gamma_{\rm E} \approx 0.577$ is the Euler constant.

The leading-in-$Z$ radiative correction to unpolarized observables from the soft and hard functions is given by  
\begin{align}\label{eq:MHspinsum}
\big\langle  |{\cal M}_H|^2 \big\rangle 
= F(Z,E)\big|_{r_H} \times \frac{4\g}{(1+\g)^2}
\,,
\end{align}
where we define $r_H^{-1}=\mu_H \e^{\gamma_{\rm E}}$. The angle brackets denote contraction
with lepton spinors, 
${\cal M}_H \to \bar{e} {\cal M}_H \gamma^0 \nu_L$, sum over final state spins, and division by the same expression in the absence of Coulomb corrections. 
Note that there is a finite multiplicative correction 
relating the  $\overline{\rm MS}$ hard function to $F(Z,E)$. 
~\\

\noindent{\bf Effective operators and anomalous dimension.}
The structure-dependent factor ${\cal M}_{\rm UV}$ appearing in Eq.~(\ref{eq:DCfac}) 
depends on the process of interest.
Important examples are beta decay transitions $[A,Z] \to [A,Z+1] e^- \bar{\nu}_e$
or $[A,Z+1] \to [A,Z] e^+ \nu_e$.
Superallowed beta
decays are governed by an EFT consisting of QED for electrons, and heavy charged scalar fields~\cite{Caswell:1985ui,Georgi:1990um,Manohar:2000dt,Paz:2015uga},
\begin{align}\label{eq:superallowedL}
 {\cal L}_{\rm eff} &=  -{\cal C} (\phi_v^{[A,Z+1]})^* \phi_v^{[A,Z]} \bar{e} 
 \slashed{v}
 (1-\gamma_5) \nu_e + {\rm H.c.} \,,
\end{align}
where $\phi_v^{[A,Z]}$ denotes a heavy scalar with 
electric charge $Z$ whose momentum fluctuations are
expanded about $p^\mu = M_{[A,Z]}v^\mu$, 
with $v^\mu=(1,0,0,0)$ in the nuclear rest frame.  
For neutron decay, the EFT involves spin-1/2 heavy fields~\cite{Caswell:1985ui,Georgi:1990um,Manohar:2000dt,Paz:2015uga},
\begin{align}\label{eq:neutronL}
    \begin{split}
  {\cal L}_{\rm eff} &= -\bar{h}_v^{(p)}\left( {\cal C}_V \gamma^\mu +{\cal C}_A \gamma^\mu \gamma_5\right) h_v^{(n)}
  \bar{e} \gamma^\mu (1-\gamma_5) \nu_e \\
  &\hspace{0.7\linewidth}+ {\rm H.c.} \,,
  \end{split}
\end{align}
where $h_v^{(p)}$ and $h_v^{(n)}$ denote spin-$1/2$ 
heavy fields with 
electric charge 1 and 0, respectively.
Matching to the EFT represented by Eqs.~(\ref{eq:superallowedL}) or (\ref{eq:neutronL}),
we identify the components of (\ref{eq:DCfac}) in terms of operator coefficients and matrix elements:
${\cal M}_{\rm UV}$ is proportional to (a linear combination 
of) ${\cal C}_i$, while 
${\cal M}_H$ and ${\cal M}_S$  give the hard and soft contributions to the EFT matrix element. 
In ${\cal M}_H$, 
at each order in $\alpha$,  the leading power of $Z$ is given by the explicit expression
(\ref{eq:MHDren}), obtained from the amplitudes (\ref{eq:DCgeneral}).  
In particular, the leading-in-$Z$ anomalous dimension 
is obtained from the $\mu_H$ dependence of Eq.~(\ref{eq:MHDren}), {\it cf}. 
Eq.~(\ref{eq:property2}) below.

We may proceed to analyze the renormalization group properties 
of weak-current operators in the EFT. 
Radiative corrections enhanced by large logarithms, $L\sim \log(\Lambda_{\rm nuc.}/E)$,
are determined by the anomalous dimensions
of the operators in~(\ref{eq:superallowedL}) and (\ref{eq:neutronL}), which 
are spin-structure independent, 
i.e., $\gamma_{A} = \gamma_V = \gamma_{\mathcal{O}}$.  
Writing
\begin{equation}
\begin{split}
  \gamma_{\cal O} 
  = {{\rm d}\log{\cal C}\over d\log\mu}
&= \sum_{n=0}^\infty \sum_{i=0}^{n+1} \left(\alpha \over 4\pi\right)^{n+1} \gamma_n^{(i)}Z^{n+1-i} \\
&\equiv \gamma^{(0)}(Z\alpha) + \alpha \gamma^{(1)}(Z\alpha) + \dots \,,
\end{split}
\end{equation}
%
we note several interesting all-orders properties:

\begin{itemize}

\item
  Powers of $Z$ greater than the power of $\alpha$ do not appear~\footnote{This can be shown explicitly using heavy particle Feynman rules in Coulomb gauge \cite{eikonal_algebra}, and confirms an old theorem due to Sirlin and B\'eg~\cite{Beg:1969zu}.}.

\item
The leading series involving $(Z\alpha)^n$ sums to
\begin{align}\label{eq:property2}
  \gamma^{(0)} &= \sqrt{1-(Z\alpha)^2} - 1 \,. 
\end{align}
This result is obtained by differentiating Eq.~(\ref{eq:MHDren}) with respect to $\mu_H$.  

\item
At each order in perturbation theory, the leading and first subleading powers of $Z$ are related~\footnote{More generally, the anomalous dimension at a given order $n$ in $\alpha$ is a linear combination of powers $Z^i(1+Z)^i$, where $2i\le n$. This is a consequence of symmetries that emerge in the limit of a massless lepton~\cite{z2a3anom}.},
\begin{align}
    \label{eq:property3}
    \gamma_{2n-1}^{(1)} =  n \gamma_{2n-1}^{(0)} ~~,~~
    \gamma_{2n}^{(2)} = n \gamma_{2n}^{(1)} \quad (n\ge 1) \,. 
\end{align}

\end{itemize}
\noindent  When $Z=0$, 
   the problem reduces to a heavy-light current operator.  
   Using our new result for $\gamma_2^{(1)}=16\pi^2(6-\pi^2/3)$~\cite{z2a3anom} and property (\ref{eq:property3}), the complete result through three-loop order at arbitrary $Z$ is 
  \begin{align}\label{eq:threeloopgamma}
    \gamma_{\cal O} &= {\alpha \over 4\pi} \gamma_0^{(1)}
    + \left({\alpha \over 4\pi}\right)^2
    \left[ 
    -8\pi^2 Z(Z+1)
    + \gamma_1^{(2)}
    \right]\\
    &\hspace{0.05\linewidth}+ 
    \left({\alpha \over 4\pi}\right)^3
    \left[ 
    16 \pi^2 \, Z(Z+1) \left( 6 - {\pi^2\over 3}\right)
    + \gamma_2^{(3)}
    \right] \,, \nonumber
\end{align}
where $\gamma_{n-1}^{(n)}$, $n=1,2,3$, 
are known from the heavy quark literature~\footnote{The relevant results are
$\gamma_0^{(1)}=-3$~\cite{Sirlin:1967zza} at one-loop, 
$\gamma_1^{(2)}= -16 \zeta_2 + \frac52 + \frac{10}{3} n_e$ 
at two-loops~\cite{Ji:1991pr}, and 
$\gamma_2^{(3)} = -80 \zeta_4 -36 \zeta_3 + 64\zeta_2 -\frac{37}{2} + n_e\left( -\frac{176}{3}\zeta_3 + \frac{448}{9}\zeta_2 + \frac{470}{9} \right) + \frac{140}{27} n_e^2$ at three-loops~\cite{Chetyrkin:2003vi} with 
  $n_e=1$ for the number of dynamical charged leptons.  The application of $\gamma_0^{(1)}$, $\gamma_1^{(2)}$ and $\gamma_2^{(3)}$ to neutron beta decay has been discussed in Ref.~\cite{Cirigliano:2023fnz}.}. 
Our result for 
$\gamma_2^{(1)}$ 
disagrees with Ref.~\cite{Jaus:1986te},
which did not include the full set of relevant diagrams at $O(Z^2\alpha^3)$~\cite{z2a3anom}.
%
%
 Note that properties (\ref{eq:property2}) 
 and (\ref{eq:property3}) 
 also determine the anomalous dimension at 
 order $Z^4\alpha^4$ and $Z^3\alpha^4$.
\\



\begin{figure}[t!]
\centering
\includegraphics[width=\linewidth]{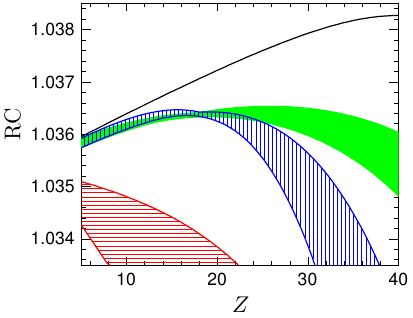}
\caption{Radiative correction to the beta decay rate as a function of nuclear charge, normalized to leading Fermi function.  
RC denotes the outer radiative correction, 
$1+\delta_R^\prime$, computed for fixed electron energy, {\it cf}. Eq.~(11) 
of Supplemental Material. 
Red, blue and green curves show results correct through resummed order $\alpha^\frac12$, $\alpha$ and $\alpha^\frac32$ respectively.  
The black curve represents the central value for 
Sirlin's heuristic estimate as implemented in
Ref.~\cite{Hardy:2020qwl}.
%
Illustrative values 
$E= 2\,{\rm MeV}$, 
$E_m = 5\,{\rm MeV}$, 
$\Lambda = 100\,{\rm MeV}$ are used for
the electron energy, 
maximum electron energy
(which enters the one-loop matrix element~\cite{Sirlin:1967zza}), 
and renormalization scale $\mu_H = \Lambda$, respectively.  
The width of the curves is given by varying $m_e/2 < \mu_L < 2 E_m$. Analytic expressions can be obtained using Eq.~(\ref{eq:resummed_interm})~\cite{supp_mat_2}. \label{fig:resum} \vspace{-12pt} }

\end{figure}

\noindent{\bf Renormalization group analysis.} 
An important advantage of identifying  
the Fermi function as a field theory object is 
the ability to resum large logarithms, 
$\sim\log(\Lambda_{\rm nuc.}/E)$, 
at high perturbative orders using renormalization group methods. 
%
%
Consider the solution to the renormalization group equation
\begin{align}
\dd \log{\cal C} = {\gamma(\alpha) \over \beta(\alpha)} \dd\alpha \,, 
\end{align}
where $\alpha$ is the ${\overline{\rm MS}}$ QED coupling (for one dynamical electron flavor)
and $\beta = \dd\alpha/\dd\log\mu = -2\alpha[ \beta_0 \alpha/(4\pi) + \beta_1 \alpha^2/(4\pi)^2 + \dots ]$~\footnote{Analytic results for the QED beta function are available through five loops~\cite{Baikov:2012zm,Herzog:2017ohr}. For our purposes we require $\beta_0=-(4/3) n_f$ and 
$\beta_1=-4 n_f$ (for $n_f=1$), {\it cf.} Ref.~\cite{Hill:2016gdf}.
}. 
Expanding $\gamma$ and $\beta$ in powers of $\alpha$ and $Z$, then integrating, 
we obtain a systematic expansion for the ratio of the renormalized operator coefficient at different scales, $C(\mu_H)/C(\mu_L)$.  Setting $\mu_H \sim \Lambda$ and $\mu_L \sim m$, we thus resum large logarithms $\log(\Lambda/m)$
\footnote{A conventional choice for $\Lambda_{\rm nuc.}$ is given in terms of the nuclear RMS charge radius, 
$\Lambda \to \sqrt{6}/R$~\cite{Sirlin:1986hpu, Hardy:2004id,Hayen:2017pwg}.
For numerical illustrations, see the Supplemental Material. 
}.
 Since the convergence of the series in $\alpha$ is influenced by $Z$, let us consider several regimes of $Z$:

\begin{itemize}

\item {\it Large $Z$ asymptotics.}

Consider a large $Z$ nucleus, counting $\log^2(\Lambda/m)  \sim \alpha^{-1}$ and $Z \sim \alpha^{-1}$. For example we may consider beta decays of $^{210}{\rm Pb}$ or $^{239}{\rm U}$. Through $O(\alpha^{1/2})$,

\begin{align}\label{eq:ClargeZ}
    &\log\qty({C(\mu_L)\over C(\mu_H)})= \nonumber\\
    ~~~~&\bigg[-\gamma^{(0)}(Z\alpha_L)  L \bigg]
    + \bigg[ b_0 \alpha_L L^2 {(Z\alpha_L)^2 \over 2\sqrt{1-(Z\alpha_L)^2}}  \bigg] ~\\[3pt]
    ~~~~~&
    + \bigg[ b_0^2 \alpha_L^2 L^3 {(Z\alpha_L)^2(3-2(Z\alpha_L)^2) \over 6(1-(Z\alpha_L)^2)^\frac32}
    - \alpha_L L \gamma^{(1)}(Z\alpha_L) 
    \bigg] \,, \nonumber\\[-6pt]
    \nonumber
\end{align}
where $\alpha_{H,L} \equiv \alpha(\mu_{H,L})$, $L=\log(\mu_H/\mu_L)$, and $b_0=-\beta_0/(2\pi)$. 
%
Consider separately the terms in $\gamma^{(1)}$ 
with odd and even powers of $(Z\alpha)$. 
Using Eq.~(\ref{eq:property3}), 
\begin{align}
    \gamma^{(1)}_{\rm odd}
    = 
    \frac12 {\partial  \over \partial(Z\alpha)} \gamma^{(0)}
    = \frac{- Z\alpha}{2\sqrt{1-(Z\alpha)^2}}
    \,.
\end{align}
The corresponding decay rate corrections involve 
(less the known $Z\alpha^2$ correction)%
\footnote{This exact result replaces the ansatz of Wilkinson~\cite{WILKINSON1997275,Hayen:2017pwg}, with differences beginning at order $Z^3\alpha^4$.}
\begin{equation}
\begin{split}
    &\delta { |C(\mu_L)|^2 \over |C(\mu_H)|^2} 
    -\alpha (Z\alpha) \log{\Lambda\over E} \\
    &= \alpha \log{\Lambda\over E} \bigg[ \frac12 (Z\alpha)^3 
    + \frac38 (Z\alpha)^5 + \dots \bigg] \,.
\end{split}
\end{equation}
The even series, $\gamma^{(1)}_{\rm even}$, is determined through three loop order by Eq.~(\ref{eq:threeloopgamma}). 
\vfill 
\pagebreak

\item {\it Intermediate $Z$.}

  Consider a medium $Z$ nucleus, counting $\log^2(\Lambda/m) \sim Z^2 \sim \alpha^{-1}$.
This is relevant for superallowed beta decays contributing to $|V_{ud}|$ extraction, which range from $Z=6$ ($^{10}{\rm C})$ to $Z=37$ ($^{74}{\rm Rb}$). 
%
%
  Through $O(\alpha^{3/2})$, the scale dependence is 
  \vspace{6pt}
  \vfill
  \begin{widetext}
  \begin{align}\label{eq:resummed_interm}
    \log\qty({C(\mu_L)\over C(\mu_H)})&= 
     {\gamma_0^{(1)} \over 2\beta_0}
    \Bigg\{
    \bigg[ \log{a_H\over a_L} + {Z^2 \gamma_1^{(0)} \over \gamma_0^{(1)} } \left( a_H - a_L \right) \bigg]
    + \bigg[
 {Z \gamma_1^{(1)} \over \gamma_0^{(1)}} (a_H-a_L) 
 \bigg]
    \\
    &\quad
    + \bigg[ \left( {\gamma_1^{(2)}\over \gamma_0^{(1)}} - {\beta_1\over\beta_0} \right)(a_H-a_L)
      + \left( {Z^2 \gamma_2^{(1)} \over \gamma_0^{(1)}} - {\beta_1\over\beta_0}{Z^2 \gamma_1^{(0)} \over \gamma_0^{(1)} } \right)\frac12 (a_H^2-a_L^2)
      + {Z^4 \gamma_3^{(0)} \over\gamma_0^{(1)}}\frac13(a_H^3-a_L^3)
      \bigg] 
    \Bigg\} \,, \nonumber
  \end{align}
\end{widetext}
where $a_{H,L} = \alpha(\mu_{H,L})/(4\pi)$ and the square brackets account for effects at order $\alpha^{\frac12}$, $\alpha^1$, $\alpha^{\frac32}$, etc.

%
Achieving permille precision demands proper treatment of terms through resummed order $\alpha^\frac32$.
This result (\ref{eq:resummed_interm})
replaces (and disagrees with) 
logarithmically enhanced contributions at order $Z^2\alpha^3$
in the ``heuristic estimate" of Sirlin and Zucchini%
\footnote{The heuristic estimate~\cite{Sirlin:1986cc,Sirlin:1986hpu} uses results from Ref.~\cite{Jaus:1972hua,Jaus:1986te}
and is defined relative to the 
contributions contained in $F(Z,E)(1+\delta_1 + \delta_2)$, where 
$F(Z,E)$ is the classical Fermi function, $\delta_1$ determines the order $\alpha$ correction, and $\delta_2$ determines the $Z\alpha^2$ correction. 
}.
Using our new result for $\gamma_2^{(1)}$~\cite{z2a3anom} we compare to this heuristic estimate, and investigate the convergence of perturbation theory in Fig.~\ref{fig:resum}.
Here we fix $\mu_H$, and 
compute the product of $|C(\mu_L)/C(\mu_H)|^2$ and the squared operator matrix element at $\mu_L$, varying $\mu_L$ as an estimate of perturbative uncertainty 
\footnote{We include the full ${\cal O}(\alpha)$ matrix element~\cite{Sirlin:1967zza} (translated to $\overline{\rm MS}$ in heavy particle effective theory). For this comparison we include only the scale-dependent logarithm, $\log(\mu/m_e)$,
in the $\alpha(Z\alpha)$ matrix element~\cite{Sirlin:1986cc}.}.
Normalizing to the leading Fermi function (known analytically to all orders) this quantity corresponds to the outer radiative correction appearing in the beta decay literature, ({\it cf}. Eq.~(11) of Supplemental Material). 
We note that 
Eq.~(\ref{eq:threeloopgamma}) is
in fact sufficient for a resummation of $C(\mu_H)/C(\mu_L)$ through $O(\alpha^2)$, although for practical applications one would also need currently unknown operator matrix elements at $O(Z\alpha^2)$~\footnote{These matrix elements have been calculated in Ref.~\cite{Sirlin:1986cc}, but not in the point-like effective theory studied here.}.
  
\item {\it Neutron beta decay.}

  Neutron beta decay corresponds to the case $Z=0$ (in our convention); we therefore define $\gamma_{n-1}\equiv\gamma_{n-1}^{(n)}$. Again counting $\log^2(\Lambda/m) \sim \alpha^{-1}$, the resummation 
  is~\footnote{
The two-loop anomalous dimension $\gamma_1$ and other subleading two-loop corrections were omitted in the leading-logarithm analysis of Ref.~\cite{Czarnecki:2004cw}.  Ref.~\cite{Cirigliano:2023fnz} included 
$\gamma_1$ in the RG flow between nucleon and electron-mass scales. 
}
\begin{equation}
\begin{split}\label{eq:resummed_neutron}
    ~~~~~&\log\qty({C(\mu_L)\over C(\mu_H)})=\\ 
    &~~~{\gamma_0 \over 2\beta_0}
    \Bigg\{
     \log{a_H\over a_L} 
     + \left( {\gamma_1\over \gamma_0} - {\beta_1\over\beta_0} \right)(a_H-a_L)
         \Bigg\} \,,
  \end{split}
\end{equation}
where the first term is of order $\alpha^\frac12$, and the second term is of order $\alpha^\frac32$. 
The complete result, correct through order $\alpha^{\frac32}$, is obtained using (\ref{eq:resummed_neutron}) together with the one-loop low-energy matrix element.
%
%
%
%

\noindent 
Even after resumming logarithms in the ratio of hadronic and electron mass scales, $\log(\Lambda/m)$, large coefficients remain in the perturbative expansion of the hard matrix element.  
While the class of
amplitudes summed in the Fermi function are enhanced at small $\beta$ and
large $Z$, neither limit holds for neutron beta decay~\footnote{At tree level, less than $0.1\%$ of the total decay rate involves electron velocity $\beta <0.1$, and less than $10\%$ of the rate involves $\beta<0.5$. }.
The large coefficients 
can instead be traced to an analytic continuation of the decay amplitude from spacelike to timelike values of momentum transfers.  The enhancements are systematically resummed by renormalization of the hard factor ${\cal M}_H$ in the factorization formula (\ref{eq:DCfac}) from negative to positive values of $\mu_S^2$ ({\it cf.} Refs.~\cite{Ahrens:2008qu,Ahrens:2009cxz}), with the result~\cite{largepi}  
%
%
\begin{align}\label{eq:MHlargepi}
       | {\cal M}_H(\mu_{S+}^2)|^2
    &= 
    \exp\bigg[{ \pi \alpha \over \beta }   \bigg]
    | {\cal M}_H(\mu_{S-}^2) |^2\,,
\end{align}
where $\mu_{S\pm}^2=\pm4p^2-\iu 0$ and 
${\cal M}_H(\mu_{S-}^2)$ is free of $\pi$-enhancements. 
%
This analysis systematically resums $\pi$-enhanced contributions, and does not rely on a non-relativistic approximation.  
%

\end{itemize}


\vfill

\noindent{\bf Discussion.}
At the outset of our discussion we posed three questions, which are now answered: 1) The scale $r^{-1}$ appearing in the Fermi function (\ref{eq:fermi})
is unambiguously related to a conventional $\overline{\rm MS}$ subtraction point, {\it cf}. Eq.~(\ref{eq:MHspinsum}). 
2) 
The Fermi function is identified as the leading-in-$Z$ contribution to 
the matrix element from the effective Lagrangian (\ref{eq:superallowedL}). 
Other radiative corrections are systematically computed using the same 
Lagrangian.  
%
%
3) 
Numerically enhanced contributions in neutron beta decay arise from perturbative logarithms $|\log[(-\vb{p}^2 -\iu 0)/(\vb{p}^2+\iu 0)]| =\pi$, and can be resummed to all orders.  
The result 
(\ref{eq:MHlargepi}) 
differs from the nonrelativistic Fermi function ansatz~\cite{Wilkinson:1982hu,Cirigliano:2023fnz} beginning at two loop order. 

%


\vfill 

Our EFT analysis allows us to systematically resum large perturbative logarithms, and to incorporate corrections that are suppressed by $1/Z$ or $E/\Lambda$. New results include: 
\vfill 
\pagebreak
\begin{enumerate}

\item

New coefficients in the expansion of the anomalous dimension for beta decay operators. We have computed the order $Z^2\alpha^3$ coefficient for the first time~\footnote{Existing order $Z^2\alpha^3$calculations~\cite{Jaus:1972hua} included only a subset of diagrams and are therefore incomplete estimates.}, and found a new symmetry linking leading-$Z$ and subleading-$Z$ terms in the perturbative expansion. Using our new result, and the existing HQET literature, we show that the first unknown coefficient occurs at 
four loops, at order $Z^2\alpha^4$~\cite{z2a3anom}. 

\item 
New 
results for the large-$Z$ asymptotics of QED radiative corrections to beta decay. We supply the infinite series of terms of order $\alpha (Z\alpha)^{2n+1} \log(\Lambda/E)$, replacing Wilkinson's ansatz~\cite{WILKINSON1997275}, and present a new result for the term of order $\alpha (Z\alpha)^2 \log(\Lambda/E)$, replacing Sirlin's heuristic estimate~\cite{Sirlin:1986cc}. 
We provide the EFT matrix element to all orders in $Z\alpha$ 
and clarify its relation to the historically employed Fermi function~\cite{Hill:2023bfh}.

\item 

An all-orders resummation of ``$\pi$-enhanced" terms in neutron beta decay, replacing the Fermi function ansatz.  This substantially improves the convergence of perturbation theory, and is important for modern applications to neutron beta decay~\cite{largepi}. 

\end{enumerate}

\noindent Each of these results has important implications 
for ongoing and near-term precision beta decay programs~\cite{UCNA:2017obv,Darius:2017arh,Fry:2018kvq,UCNt:2021pcg,Shidling:2014ura,Eibach:2015ksa,Sternberg:2015nnr,Gulyuz:2016ppg,Fenker:2016mka,Long:2017gdh,Fenker:2017rcx,Brodeur:2016cci,Shidling:2018fvb,Valverde:2018haz,OMalley:2020vop,Burdette:2020bke,Long:2020lby,Muller:2022jew,Long:2022yea}. 
Detailed computations are presented elsewhere~\cite{z2a3anom,Hill:2023bfh,largepi}. Related work on new eikonal identities for charged current processes is presented in Ref.~\cite{eikonal_algebra}. 
The same formalism applies to any situation involving charged leptons and nuclei, provided the lepton energy is small compared to the inverse nuclear radius.

An immediate conclusion of our study is that the existing estimate for $O(Z^2\alpha^3)$ corrections is incorrect.
%
%
Focusing on the dominant logarithmically enhanced terms, 
the coefficient ``$a$'' in Sirlin's heuristic estimate~\cite{Sirlin:1986cc,Jaus:1986te}, changes. 
For the 9 transitions with smallest ${\cal F}t$ uncertainty (at or below permille level),
this leads to shifts ranging from 
$1.1\times 10^{-4}$ for $^{14}{\rm O}$ to 
$1.4 \times 10^{-3}$ for $^{54}{\rm Co}$~\footnote{See Supplemental Material [URL-HERE] for a discussion of how these numerical shifts are computed.}
i.e., an order of magnitude larger than the estimated uncertainty  on the outer radiative correction \cite{Hardy:2020qwl}. 
We observe that these shifts are comparable in magnitude 
to the CKM discrepancy, $|V_{ud}|^2 + |V_{us}|^2 + |V_{ub}|^2 - 1 = -0.0015(6)$~\cite{Hardy:2020qwl}, and with a sign that goes in the direction of resolving the discrepancy.  
Accounting for these strongly $Z$-dependent corrections should 
also impact New Physics constraints such as on 
scalar currents beyond the Standard Model~\cite{Hardy:2020qwl}. 
%
%
A complete determination of the long-distance radiative corrections at the $10^{-4}$ level will require revisiting the $O(Z\alpha^2)$ matrix element in the point-like EFT considered here; this work is ongoing. Future work will address factorization at subleading power, and investigate the impact on phenomenology including hadronic~\cite{Ando:2004rk,Cirigliano:2022hob}
and nuclear~\cite{Seng:2018qru,Gennari:INT} matching uncertainties.

\vfill
\pagebreak

\begin{acknowledgments}\noindent{\bf Acknowledgments.}  We thank Susan Gardner and Oleksandr Tomalak for useful discussions regarding radiative corrections for beta decays, and Peter Vander Griend for collaboration on Ref.~\cite{largepi}. RP thanks the Institute for Nuclear Theory at the University of Washington for its kind hospitality and stimulating research environment during program INT 23-1B. This research was supported in part by the INT's U.S. Department of Energy grant No. DE-FG02-00ER41132. This work was
supported by the U.S. Department of Energy, Office of Science, Office
of High Energy Physics, under Award DE-SC0019095.
Fermilab is operated by Fermi Research Alliance, LLC
under Contract No. DE-AC02-07CH11359 with the United States Department
of Energy. RP is supported by the Neutrino Theory Network under Award Number DEAC02-07CHI11359, the U.S. Department of Energy, Office of Science, Office of High Energy Physics, under Award Number DE-SC0011632, and by the Walter Burke Institute for Theoretical Physics.   
RJH gratefully acknowledges support from the Institute for Advanced Study, where a part of this work was completed.
\end{acknowledgments}

\pagebreak

\onecolumngrid 
\section{Supplemental material}

\section{Factorization}

\begin{figure}[htb]
\begin{center}

\vspace{5mm}

\parbox{25mm}{
\begin{fmfgraph*}(60,40)
\fmfstraight
\fmftopn{t}{3}
\fmfbottomn{b}{3}
\fmf{double}{b1,v,b3}
\fmf{fermion}{t1,v,t3}
\fmfv{decor.shape=circle,decor.filled=shaded}{v}
\fmflabel{$\psi_v^\prime$}{b3}
\fmflabel{$\psi_v$}{b1}
\fmflabel{$e$}{t3}
\fmflabel{$\nu_e$}{t1}
\end{fmfgraph*}
} = $-i \Gamma_{\rm UV} \otimes \Gamma_e$
\qquad
\parbox{10mm}{
\begin{fmfgraph*}(30,30)
    \fmfstraight
    \fmftop{t}
    \fmfbottomn{b}{3}
    \fmf{photon}{t,b2}
    \fmf{double}{b1,b2,b3}
    \fmflabel{$\mu$}{t}
    \fmflabel{$\psi_v^\prime$}{b1}
\end{fmfgraph*}
}
= $i (Z+1)e \, v^\mu$ 
\qquad
\parbox{10mm}{
\begin{fmfgraph*}(30,30)
    \fmfstraight
    \fmftop{t}
    \fmfbottomn{b}{3}
    \fmf{photon}{t,b2}
    \fmf{double}{b1,b2,b3}
    \fmflabel{$\mu$}{t}
    \fmflabel{$\psi_v$}{b1}
\end{fmfgraph*}
}
= $i Ze \, v^\mu$ 
\qquad
\parbox{10mm}{
\begin{fmfgraph*}(30,30)
    \fmfstraight
    \fmftop{t}
    \fmfbottomn{b}{3}
    \fmf{photon}{t,b2}
    \fmf{fermion}{b1,b2,b3}
    \fmflabel{$\mu$}{t}
    \fmflabel{$e$}{b1}
\end{fmfgraph*}
}
= $-i e \, \gamma^\mu$ 
\vspace{10mm}

\parbox{28mm}{
\begin{fmfgraph*}(60,40)
  \fmfleftn{l}{3}
  \fmfrightn{r}{3}
  \fmf{photon}{l2,x,y,r2}
  \marrow{a}{up}{top}{$q$}{x,y}
  \fmflabel{$\mu$}{l2}
  \fmflabel{$\nu$}{r2}
\end{fmfgraph*}
}\,\,
= ${\displaystyle {-i \over q^2 -\lambda^2 + i0} \left( g_{\mu\nu} - (1-\xi) {q_\mu q_\nu \over q^2}  \right)}$
\vspace{3mm}

\parbox{28mm}{
\begin{fmfgraph*}(60,40)
  \fmfleftn{l}{3}
  \fmfrightn{r}{3}
  \fmf{double}{l2,x,y,r2}
  \marrow{a}{up}{top}{$q$}{x,y}
  \fmflabel{$\psi_v$}{l2}
\end{fmfgraph*}
}
= ${\displaystyle {i \over v\cdot q + i0}}$
\qquad
\parbox{28mm}{
\begin{fmfgraph*}(60,40)
  \fmfleftn{l}{3}
  \fmfrightn{r}{3}
  \fmf{double}{l2,x,y,r2}
  \marrow{a}{up}{top}{$q$}{x,y}
  \fmflabel{$\psi_v^\prime$}{l2}
\end{fmfgraph*}
}
= ${\displaystyle {i \over v\cdot q + i0}}$
\qquad
\parbox{28mm}{
\begin{fmfgraph*}(60,40)
  \fmfleftn{l}{3}
  \fmfrightn{r}{3}
  \fmf{phantom}{l2,x,y,r2}
  \fmf{fermion}{l2,r2}
    \noarrow{a}{up}{top}{$q$}{x,y}
  \fmflabel{$e$}{l2}
\end{fmfgraph*}
}
= ${\displaystyle {i \over \slashed{q} -m + i0}}$

\end{center}
\caption{\label{fig:FRH} Feynman rules for the hard-scale Lagrangian (\ref{eq:LH}), with gauge parameter $\xi$ and photon mass $\lambda$}
\end{figure}
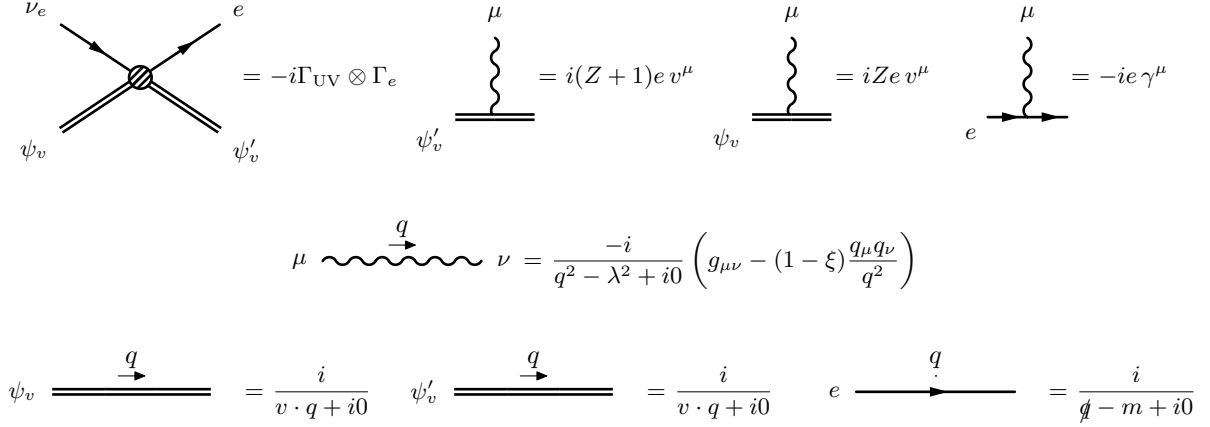

Consider the Lagrangian (Eq.~(6) and Eq.~(7) of the main text) 
describing physics below hadronic and nuclear scales: 
\begin{align}\label{eq:LH}
{\cal L}_{H} &=  -\bar{\psi}_v^\prime \Gamma_{\rm UV} \psi_v 
\otimes 
\bar{e} \,\Gamma_e \nu_e + {\rm H.c.} \,, 
\end{align}
where $v^\mu= (1,0,0,0)$ in the nuclear rest frame.  
For the superallowed beta decay case we have scalar heavy particle fields 
$\psi_v= \phi^{[A,Z]}_v$, $\psi_v^\prime = \phi^{[A,Z+1]}_v$, and Dirac structures  
$\Gamma_{\rm UV} = {\cal C}$ and $\Gamma_e = \slashed{v} (1-\gamma_5)$. 
For the neutron beta decay case we have fermionic heavy particle fields 
$\psi_v= h^{(n)}_v$, $\psi_v^\prime = h_v^{(p)}$, and Dirac structures  
$\Gamma_{\rm UV} = {\cal C}_V \gamma^\mu + {\cal C}_A \gamma^\mu \gamma_5$ 
and $\Gamma_e = \gamma_\mu (1-\gamma_5)$. 
Coefficients ${\cal C}$, ${\cal C}_V$ and ${\cal C}_A$ are determined by matching the effective theory (\ref{eq:LH}) to the quark-level Standard Model Lagrangian.  Feynman rules corresponding to this Lagrangian are shown in Fig.~\ref{fig:FRH}.

\begin{figure}[htb]
\begin{center}

\vspace{5mm}

\parbox{25mm}{
\begin{fmfgraph*}(60,40)
\fmfstraight
\fmftopn{t}{3}
\fmfbottomn{b}{3}
\fmf{double}{b1,v,b3}
\fmf{fermion}{t1,v}
\fmf{double}{v,t3}
\fmfv{decor.shape=circle,decor.filled=shaded}{v}
\fmflabel{$\psi_v^\prime$}{b3}
\fmflabel{$\psi_v$}{b1}
\fmflabel{$h_{v^\prime}^{(e)}$}{t3}
\fmflabel{$\nu_e$}{t1}
\end{fmfgraph*}
} = $-i \Gamma_{\rm UV} \otimes \bigg[ A(v\cdot v^\prime) + B(v\cdot v^\prime )\slashed{v} \bigg] \Gamma_e$
\qquad\qquad
\parbox{10mm}{
\begin{fmfgraph*}(30,30)
    \fmfstraight
    \fmftop{t}
    \fmfbottomn{b}{3}
    \fmf{photon}{t,b2}
    \fmf{fermion}{b1,b2,b3}
    \fmflabel{$\mu$}{t}
    \fmflabel{$h_{v^\prime}^{(e)}$}{b1}
\end{fmfgraph*}
}
= $-i e \, v^{\prime \mu}$ 
\vspace{10mm}

\parbox{28mm}{
\begin{fmfgraph*}(60,40)
  \fmfleftn{l}{3}
  \fmfrightn{r}{3}
  \fmf{double}{l2,x,y,r2}
  \marrow{a}{up}{top}{$q$}{x,y}
  \fmflabel{$h_{v^\prime}^{(e)}$}{l2}
\end{fmfgraph*}
}
= ${\displaystyle {i \over v^\prime\cdot q + i0}}$

\end{center}
\caption{\label{fig:FRS} Feynman rules for the soft-scale Lagrangian (\ref{eq:LS})
involving the heavy electron field.}
\end{figure}
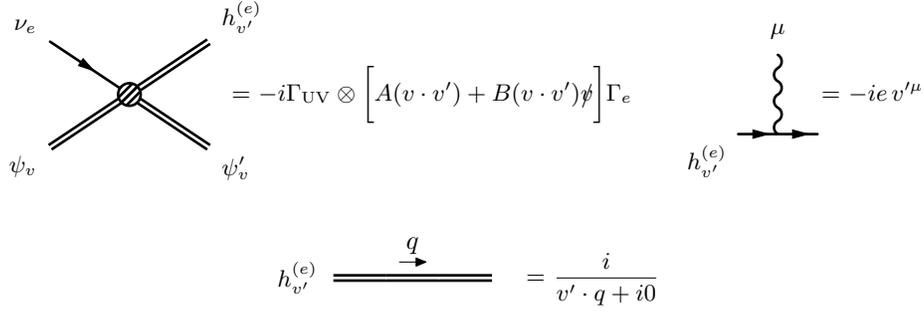

For physics below the electron mass scale, 
electrons of a given four velocity $v^\prime_\mu = p^\mu/m$ (plus soft radiation) are described by a Lagrangian where the electron is also represented by a heavy-particle field 
\footnote{
Recall that we are counting $v\cdot v^\prime = E/m$ as order unity.  
Since the electron is pointlike there is no further restriction on $v\cdot v^\prime$, unlike the case for the decay of heavy mesons with internal structure~\cite{Isgur:1990yhj}. 
},
\begin{align}\label{eq:LS}
    {\cal L}_{S} &=  -\bar{\psi}_v^\prime \Gamma_{\rm UV} \psi_v 
\otimes 
\bar{h}^{(e)}_{v^\prime}\bigg[ A(v\cdot v^\prime) + B(v\cdot v^\prime )\slashed{v} \bigg]  \,\Gamma_e \nu_e + {\rm H.c.} \,. 
\end{align}
Feynman rules corresponding to this Lagrangian are shown in Fig.~\ref{fig:FRS}.
It is readily seen that the heavy particle effective theory Feynman rules ensure that the most general Dirac structure is given by the square bracket in Eq.~(\ref{eq:LS}) (note that we have $\slashed{v}^\prime h_{v}^\prime = h_{v}^\prime$).  
From the soft Lagrangian (\ref{eq:LS}) we may read off the complete amplitude in the 
factorized form, Eq.~(3) of the main text.  Writing the complete amplitude as 
\begin{align}
\bar{u}^{(\psi^\prime)} \Gamma_{\rm UV} u^{(\psi)}  \otimes 
 \bar{u}^{(e)}(p) {\cal M}_S {\cal M}_H \Gamma_e v^{(\nu_e)}
\end{align}
we identify 
\begin{align}
    {\cal M}_{\rm UV} &=  \bar{u}^{(\psi^\prime)} \Gamma_{\rm UV} u^{(\psi)} \,, 
    \\ \nonumber
    {\cal M}_H &= A(v\cdot v^\prime) + B(v\cdot v^\prime )\slashed{v} \,,
\end{align}
and the soft matrix element is given by 
\begin{align}
\left\langle 
    \bar{\psi}_v^\prime \Gamma_{\rm UV} \psi_v \, \bar{h}^{(e)}_{v^\prime} \right\rangle 
= 
      {\cal M}_S \, \left\langle \bar{\psi}_v^\prime \Gamma_{\rm UV} \psi_v \, \bar{h}^{(e)}_{v^\prime} \right\rangle_{\rm tree}
      = 
\bar{u}^{(\psi^\prime)} \Gamma_{\rm UV} u^{(\psi)} 
    \bar{u}^{(e)}(p) {\cal M}_S 
    \,.
\end{align}

\vfill 
\pagebreak

\section{Real radiation}
Real photon radiation does not contribute to the leading-in-$Z$ 
radiative correction described by the Fermi function ({\it cf}. Eq.~(1) of the main text), but does contribute at subleading orders.  
The Lagrangian (\ref{eq:LH}) describes physics below hadronic and nuclear scales, including arbitrary photon radiation.  In particular, the solution 
to the renormalization group equation, Eq.~(12) of the main text ({\it cf}. Eqs.~(13), (16) of the main text), applies to both processes with and without real radiation, and systematically resums large logarithms $\sim \log(\Lambda/m)$.

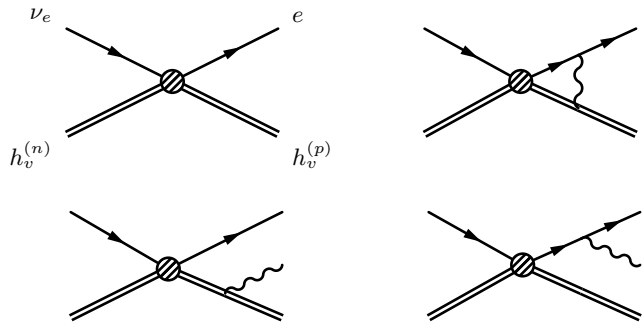
\begin{figure}[htb]
\begin{center}

\vspace{5mm}

\parbox{40mm}{
\begin{fmfgraph*}(80,40)
\fmfstraight
\fmftopn{t}{3}
\fmfbottomn{b}{3}
\fmf{double}{b1,v,b3}
\fmf{fermion}{t1,v,t3}
\fmfv{decor.shape=circle,decor.filled=shaded}{v}
\fmflabel{$h_v^{(p)}$}{b3}
\fmflabel{$h_v^{(n)}$}{b1}
\fmflabel{$e$}{t3}
\fmflabel{$\nu_e$}{t1}
\end{fmfgraph*}
}
\qquad
\parbox{40mm}{
\begin{fmfgraph*}(80,40)
\fmfstraight
\fmftopn{t}{3}
\fmfbottomn{b}{3}
\fmf{double,tension=0.6}{b1,v}
\fmf{double}{v,w1,b3}
\fmf{fermion,tension=0.6}{t1,v}
\fmf{fermion}{v,w2,t3}
\fmffreeze
\fmf{photon,tension=0.5}{w1,w2}
\fmfv{decor.shape=circle,decor.filled=shaded}{v}
\end{fmfgraph*}
}
\vspace{10mm}

\parbox{40mm}{
\begin{fmfgraph*}(80,40)
\fmfstraight
\fmftopn{t}{3}
\fmfbottomn{b}{3}
\fmfrightn{r}{3}
\fmf{double,tension=0.75}{b1,v}
\fmf{double,tension=2}{v,w1,b3}
\fmf{phantom}{t1,v,w2,t3}
\fmffreeze
\fmf{fermion,tension=0.5}{t1,v}
\fmf{fermion,tension=2}{v,t3}
\fmf{photon}{w1,r2}
\fmfv{decor.shape=circle,decor.filled=shaded}{v}
\end{fmfgraph*}
}
\qquad
\parbox{40mm}{
\begin{fmfgraph*}(80,40)
\fmfstraight
\fmftopn{t}{3}
\fmfbottomn{b}{3}
\fmfrightn{r}{3}
\fmf{double,tension=1.5}{b1,v,w1,b3}
\fmf{fermion}{t1,v}
\fmf{fermion,tension=2.5}{v,w2,t3}
\fmffreeze
\fmf{photon}{r2,w2}
\fmfv{decor.shape=circle,decor.filled=shaded}{v}
\end{fmfgraph*}
}

\end{center}
\caption{\label{fig:Sirlin} Feynman diagrams for first order corrections to neutron beta decay rate computed from Eq.~(\ref{eq:LH}).  Wavefunction renormalization is not shown.}
\end{figure}

To demonstrate the application of (\ref{eq:LH}) to real photon radiation, 
let us consider the leading, order $\alpha$, correction, computed from the diagrams shown in Fig.~\ref{fig:Sirlin}. 
Let us write 
\begin{align}\label{eq:Sirlin}
    \sigma = \sigma_{\rm tree}(\mu_H) \bigg[ 1 + {\alpha\over 2\pi}\bigg( \Delta_{S,V} + \Delta_{S,R} + \Delta_{H,V} + \Delta_{H,R}\bigg) + \dots \bigg] \,.
\end{align}
The virtual and real soft contributions are given by Feynman diagrams in the soft theory, {\it cf}. (\ref{eq:LS}) and Fig.~\ref{fig:FRS}:
\begin{align}
    \Delta_{S,V} &= \log{\mu_S\over\lambda}\left( 4 -{2\over \beta}\log{1+\beta\over 1-\beta} \right) \,,
    \\
    \Delta_{S,R} &= \log{2\Delta E\over \lambda}\left( {2\over \beta}\log{1+\beta\over1-\beta}-4\right)
    - {1\over 2\beta} \log^2\left(1+\beta\over 1-\beta\right) 
    -{2\over\beta}{\rm Li}_2\left(2\beta\over 1+\beta\right) + {1\over\beta}\log{1+\beta\over 1-\beta} 
    +2 \,,
    \label{Delta_SR}
\end{align}
where real photons of energy smaller than $\Delta E$ are included (we may take $\Delta E \ll m$). 
The remaining hard contributions are 
\begin{align}
    \Delta_{H,V} &= 3\log{\mu_H\over m} + \log{\mu_S\over m}\left( {2\over \beta}\log{1+\beta\over 1-\beta} - 4\right) 
    + \beta \log{1+\beta\over 1-\beta} + {2\pi^2\over \beta} - {2\over\beta}{\rm Li}_2\left(2\beta\over 1+\beta\right) - 
\frac{1}{2\beta}\log^2\left(1+\beta\over 1-\beta\right) \,,
     \\
     \Delta_{H,R} &= \log{\Delta E \over \Em - E} \left(4 - {2\over \beta}\log{1+\beta\over 1-\beta} \right) 
     + 
\frac{1}{\beta}
\log{1+\beta\over 1-\beta}\left[ {(\Em-E)^2\over 12 E^2} 
     + {2(\Em-E)\over 3E}- 3
     \right]
     - {4(\Em-E)\over 3E} + 6 \,, 
\end{align}
where $\Em$ is the nuclear energy difference, equal to the maximum possible electron energy. 
It is readily seen that the dependence on the photon mass regulator cancels in the sum of $\Delta_{S,V}$ and $\Delta_{S,R}$, and the soft photon threshold $\Delta E$ cancels in the sum of $\Delta_S= \Delta_{S,V}+\Delta_{S,R}$ and $\Delta_H = \Delta_{H,V}+\Delta_{H,R}$.  
Further, the dependence on the arbitrary factorization scale $\mu_S$ cancels between $\Delta_S$ and $\Delta_H$, and the dependence on $\mu_H$ cancels between $\Delta_H$ and $\sigma_{\rm tree}(\mu_H)$.  For the latter cancellation, we observe that $d\log\sigma_{\rm tree}(\mu_H)/d\log\mu_H = 3\alpha/(2\pi) + \dots$, using Eq.~(11) of the main text (at $Z=0$ corresponding to neutron beta decay).

The expression (\ref{eq:Sirlin}) corresponds to the well-known result of Sirlin~\cite{Sirlin:1967zza}.  In particular, upon setting $\mu_H = m_p$ ($m_p$ the proton mass) and omitting the term $2\pi^2/\beta$ in our $\Delta_{H,V}$, the sum $\Delta = \Delta_S + \Delta_H$ gives $g(E,\Em,m)$ in Eq.~(20b) from Ref.~\cite{Sirlin:1967zza} up to a conventional constant ($+2$ in Eq.~(\ref{Delta_SR}) for $\overline{\rm MS}$ 
vs. $-3/4$ in Ref.~\cite{Sirlin:1967zza}, {\it cf.} Ref.~\cite{Ando:2004rk}). In Ref.~\cite{Sirlin:1967zza},
the additional $2\pi^2/\beta$ term is included in a Fermi function factor, 
{\it cf}.~Eq.~(1) of the main text at $Z=1$. In comparing $g(E,\Em,m)$ with $\Delta$ we have used that $L(x)=(-1)\times {\rm Li}_2(x)$.

\section{Numerical impact of radiative corrections}

Our analysis determines new contributions to structure-independent radiative corrections to superallowed nuclear beta decay rates.  These corrections directly impact the associated extraction of $|V_{ud}|$.
In the notation of Ref.~\cite{Hardy:2020qwl} we have
\begin{equation}\label{eq:deltaRp}
    1+\delta_R' \simeq  \qty[ \frac{C(\mu_L)\big/C(\mu_H)}{\exp\qty[(1-\sqrt{1-Z^2\alpha^2})~\log(\mu_H/\mu_L)]}]^2 \qty(\frac{\int \dd \Pi \, |\mathcal{M}|^2(\mu_L) }{\int \dd \Pi ~ F(Z,E)|_{r_L} \times \tfrac{4\eta}{(1+\eta)^2} }  ) \,.
\end{equation}
where ${\cal M}$ is the effective operator matrix element and 
$\int \dd \Pi$ denotes integration over (electron/positron, neutrino and real photon) 
phase space and sum over lepton spin states 
\footnote{
Remaining radiative corrections, denoted by 
$(1+\Delta_R^V)(1+\delta_{\rm NS} - \delta_C)$ in 
Ref.~\cite{Hardy:2020qwl}, 
are identified (up to a prefactor involving $G_F V_{ud}$) 
with the matching coefficient ${\cal C}(\mu_H)$ 
in the effective Lagrangian (\ref{eq:LH}).
}.
The product of denominators in Eq.~(\ref{eq:deltaRp}) extracts the conventional integrated Fermi function prefactor appearing in traditional rate formulas~\cite{Hardy:2020qwl}.  Here, as in the main text, the short distance regulator in $F(Z,E)$ is identified as $r_{H,L} \equiv \mu_{H,L}e^{\gamma_E}$. 
Taking $\mu_H \sim \Lambda$ and $\mu_L \sim m$, the round bracket in Eq.~(\ref{eq:deltaRp}) is free of large logarithms $L =\log(\Lambda/m)$.  
Further, because we have divided by the leading-in-$Z$ expression, the perturbative expansion contains only terms $Z^m \alpha^n$ with $m<n$.  In the ``intermediate $Z$" power counting discussed in the main text, the relevant terms are at order $\alpha$~\cite{Sirlin:1967zza} ({\it cf}. Eq.~(\ref{eq:Sirlin}))
and at order $Z\alpha^2 \sim \alpha^\frac32$. 
The order $Z\alpha^2$ matrix element has been considered using a different regulator scheme in Ref.~\cite{Sirlin:1986cc}. 
Further discussion of this matrix element will be presented elsewhere~\cite{largepi}. 

Logarithmic enhancements are contained in the square bracket in Eq.~(\ref{eq:deltaRp}). 
Our results for the resummed coefficient, Eq.~(16) of the main text,  yield 
\begin{multline}
    \qty[ \frac{C(\mu_L)\big/C(\mu_H)}{\exp\qty[(1-\sqrt{1-Z^2\alpha^2})~\log(\mu_H/\mu_L)]}]^2
    = 1 + \alpha \bigg[ {3\over 2\pi} L \bigg]
    + \alpha^2 \bigg[ Z L +  {13 \over 8\pi^2} L^2 + \left(\frac13 - {35\over 48\pi^2}\right) L  \bigg]
    \\
+    \alpha^3 \bigg[ {2\over 3\pi}Z^2 L^2 
+ \left( {\pi \over 6} - {3\over \pi}  \right) Z^2 L 
+ {13\over 6\pi} Z L^2 
+ {221 \over 144 \pi^3}L^3 \bigg]
+ \alpha^4 \bigg[ {13\over 9\pi^2} Z^2 L^3 \bigg]  + \dots \,,
\end{multline}
expressed in terms of the onshell coupling $\alpha$. 
The terms at order $\alpha L$ and at order $\alpha^2 Z L$ 
have been included as part of the order $\alpha$~\cite{Sirlin:1967zza} and order $Z\alpha^2$~\cite{Sirlin:1986cc} corrections. 
The term at order $\alpha^3 Z^2 L^2$ is a correction to the two-loop Fermi function, and has been included in Ref.~\cite{Sirlin:1986cc}.
The term at order $\alpha^3 Z^2 L$ corresponds to $\gamma_2^{(1)}$ and corrects a previous result in the literature, as noted in the main text.   
The remaining terms have not to our knowledge been included 
in analyses of nuclear beta decay:
the term at order $\alpha^2 L$ involves the two-loop anomalous dimension $\gamma_1^{(2)}$;
the terms at order $\alpha^2 L^2$ and $\alpha^3 L^3$ are higher-order corrections that involve the one-loop anomalous dimension $\gamma_0^{(1)}$;
the term at order $\alpha^3 Z L^2$ is a higher order correction induced by the $Z\alpha^2$ contribution (involving $\gamma_1^{(1)}$);   
and the term at order $\alpha^4 Z^2 L^3$ involves the interference of the two loop Fermi function and one loop anomalous dimension, $\gamma_1^{(0)}$ and $\gamma_0^{(1)}$. 

\begin{table}[t]
\begin{center}
\begin{tabular}{ |c|c| } 
\hline
transition & $(\Delta a)\times Z^2\alpha^3\log(\Lambda/m)$ 
\\
 \hline
$~^{14}{\rm O} \rightarrow \!~^{14}{\rm N}$     &    $-1.1\times 10^{-4}$    \\
$~^{26m}{\rm Al} \rightarrow\! ~^{26}{\rm Mg}$ & $-3.2\times 10^{-4}$    \\
$~^{34}{\rm Cl} \rightarrow \!~^{34}{\rm S}$ & $-5.6 \times 10^{-4}$ \\
$~^{34}{\rm Ar} \rightarrow \!~^{34}{\rm Cl}$ & $-6.3 \times 10^{-4}$ \\
$~^{38m}{\rm K} \rightarrow \!~^{38}{\rm Ar}$ & $-7.1 \times 10^{-4}$ \\
$~^{42}{\rm Sc} \rightarrow \!~^{42}{\rm Ca}$ & $-8.7 \times 10^{-4}$ \\
$~^{46}{\rm V} \rightarrow \!~^{46}{\rm Ti}$ & $-10.5 \times 10^{-4}$ \\
$~^{50}{\rm Mn} \rightarrow \!~^{50}{\rm Cr}$ & $-12.5 \times 10^{-4}$ \\
$~^{54}{\rm Co} \rightarrow \!~^{54}{\rm Fe}$ &      $-14.6\times 10^{-4}$ \\
 \hline
\end{tabular}
\caption{Shift in 
the outer radiative correction at order $Z^2\alpha^3 \log(\Lambda/m)$, for the 9 transitions with smallest ${\cal F}t$ uncertainty in Ref.~\cite{Hardy:2020qwl}.
\label{tab:Deltaa}}
\end{center}
\end{table}

Let us focus on the $\alpha^3 Z^2 L$ correction and 
determine the numerical impact of
 our new calculation compared to the Jaus-Rasche estimate of the anomalous dimension. 
 We use the updated value from their 1987 paper~\cite{Jaus:1986te}, which determines the coefficient $a$ in Sirlin's heuristic estimate of $\delta_3$ (i.e., $a=(\pi^2/3-3/2)/\pi \approx 0.5697$ as written in Footnote~10 of Ref.~\cite{Sirlin:1986hpu}).  We find instead 
 $a=-\gamma_{2}^{(1)}/(32\pi^3)\approx -0.4313$. 
We define $\Delta a =-0.4313 - 0.5697\approx -1.001$. 
To estimate the size of the logarithm we set $L=\log(\Lambda/m)$ with $\Lambda= \sqrt{6}/\sqrt{\langle r^2 \rangle}$ and $\langle r^2 \rangle$ taken from Ref.~\cite{Hardy:2004id}. In the convention we follow in the main text, the two nuclei involved in the transition have charge $Z$ and $Z+1$. The superallowed transitions that are important for $|V_{ud}|$ extractions are of the form $[A,Z+1] \rightarrow e^+ \nu_e [A,Z]$ and for simplicity we will always choose transitions with a positron in the final state. This means that $Z$ will always correspond to the daughter nucleus. 
For the transitions with relatively small errors~\cite{Hardy:2020qwl} we find the results in Table~\ref{tab:Deltaa} 
\footnote{
The Table lists transitions for which the total ${\cal F}t$ 
uncertainty is $\lesssim 10^{-3}$, {\it cf}. Fig.~4 of Ref.~\cite{Hardy:2020qwl}.}.
In each case, the shift is larger than the error currently ascribed to $\delta_R'$.
A definitive statement at an accuracy on the order of 100 ppm will require a renewed scrutiny of the matrix element at $O(Z\alpha^2)$ in the point-like effective theory.

\end{fmffile}

\bibliography{fermi}

\begin{thebibliography}{109}%
\makeatletter
\providecommand \@ifxundefined [1]{%
 \@ifx{#1\undefined}
}%
\providecommand \@ifnum [1]{%
 \ifnum #1\expandafter \@firstoftwo
 \else \expandafter \@secondoftwo
 \fi
}%
\providecommand \@ifx [1]{%
 \ifx #1\expandafter \@firstoftwo
 \else \expandafter \@secondoftwo
 \fi
}%
\providecommand \natexlab [1]{#1}%
\providecommand \enquote  [1]{``#1''}%
\providecommand \bibnamefont  [1]{#1}%
\providecommand \bibfnamefont [1]{#1}%
\providecommand \citenamefont [1]{#1}%
\providecommand \href@noop [0]{\@secondoftwo}%
\providecommand \href [0]{\begingroup \@sanitize@url \@href}%
\providecommand \@href[1]{\@@startlink{#1}\@@href}%
\providecommand \@@href[1]{\endgroup#1\@@endlink}%
\providecommand \@sanitize@url [0]{\catcode `\\12\catcode `\$12\catcode
  `\&12\catcode `\#12\catcode `\^12\catcode `\_12\catcode `\%12\relax}%
\providecommand \@@startlink[1]{}%
\providecommand \@@endlink[0]{}%
\providecommand \url  [0]{\begingroup\@sanitize@url \@url }%
\providecommand \@url [1]{\endgroup\@href {#1}{\urlprefix }}%
\providecommand \urlprefix  [0]{URL }%
\providecommand \Eprint [0]{\href }%
\providecommand \doibase [0]{http://dx.doi.org/}%
\providecommand \selectlanguage [0]{\@gobble}%
\providecommand \bibinfo  [0]{\@secondoftwo}%
\providecommand \bibfield  [0]{\@secondoftwo}%
\providecommand \translation [1]{[#1]}%
\providecommand \BibitemOpen [0]{}%
\providecommand \bibitemStop [0]{}%
\providecommand \bibitemNoStop [0]{.\EOS\space}%
\providecommand \EOS [0]{\spacefactor3000\relax}%
\providecommand \BibitemShut  [1]{\csname bibitem#1\endcsname}%
\let\auto@bib@innerbib\@empty
\bibitem [{\citenamefont {Nunokawa}\ \emph {et~al.}(2008)\citenamefont
  {Nunokawa}, \citenamefont {Parke},\ and\ \citenamefont
  {Valle}}]{Nunokawa:2007qh}%
  \BibitemOpen
  \bibfield  {author} {\bibinfo {author} {\bibfnamefont {H.}~\bibnamefont
  {Nunokawa}}, \bibinfo {author} {\bibfnamefont {S.~J.}\ \bibnamefont {Parke}},
  \ and\ \bibinfo {author} {\bibfnamefont {J.~W.~F.}\ \bibnamefont {Valle}},\
  }\href {\doibase 10.1016/j.ppnp.2007.10.001} {\bibfield  {journal} {\bibinfo
  {journal} {Prog. Part. Nucl. Phys.}\ }\textbf {\bibinfo {volume} {60}},\
  \bibinfo {pages} {338} (\bibinfo {year} {2008})},\ \Eprint
  {http://arxiv.org/abs/0710.0554} {arXiv:0710.0554 [hep-ph]} \BibitemShut
  {NoStop}%
\bibitem [{\citenamefont {Abe}\ \emph {et~al.}(2014)\citenamefont {Abe} \emph
  {et~al.}}]{Hyper-KamiokandeWorkingGroup:2014czz}%
  \BibitemOpen
  \bibfield  {author} {\bibinfo {author} {\bibfnamefont {K.}~\bibnamefont
  {Abe}} \emph {et~al.} (\bibinfo {collaboration} {Hyper-Kamiokande Working
  Group})\ }(\bibinfo {year} {2014})\ \Eprint {http://arxiv.org/abs/1412.4673}
  {arXiv:1412.4673 [physics.ins-det]} \BibitemShut {NoStop}%
\bibitem [{\citenamefont {Diwan}\ \emph {et~al.}(2016)\citenamefont {Diwan},
  \citenamefont {Galymov}, \citenamefont {Qian},\ and\ \citenamefont
  {Rubbia}}]{Diwan:2016gmz}%
  \BibitemOpen
  \bibfield  {author} {\bibinfo {author} {\bibfnamefont {M.~V.}\ \bibnamefont
  {Diwan}}, \bibinfo {author} {\bibfnamefont {V.}~\bibnamefont {Galymov}},
  \bibinfo {author} {\bibfnamefont {X.}~\bibnamefont {Qian}}, \ and\ \bibinfo
  {author} {\bibfnamefont {A.}~\bibnamefont {Rubbia}},\ }\href {\doibase
  10.1146/annurev-nucl-102014-021939} {\bibfield  {journal} {\bibinfo
  {journal} {Ann. Rev. Nucl. Part. Sci.}\ }\textbf {\bibinfo {volume} {66}},\
  \bibinfo {pages} {47} (\bibinfo {year} {2016})},\ \Eprint
  {http://arxiv.org/abs/1608.06237} {arXiv:1608.06237 [hep-ex]} \BibitemShut
  {NoStop}%
\bibitem [{\citenamefont {Acero}\ \emph {et~al.}(2022)\citenamefont {Acero}
  \emph {et~al.}}]{NOvA:2021nfi}%
  \BibitemOpen
  \bibfield  {author} {\bibinfo {author} {\bibfnamefont {M.~A.}\ \bibnamefont
  {Acero}} \emph {et~al.} (\bibinfo {collaboration} {NOvA}),\ }\href {\doibase
  10.1103/PhysRevD.106.032004} {\bibfield  {journal} {\bibinfo  {journal}
  {Phys. Rev. D}\ }\textbf {\bibinfo {volume} {106}},\ \bibinfo {pages}
  {032004} (\bibinfo {year} {2022})},\ \Eprint
  {http://arxiv.org/abs/2108.08219} {arXiv:2108.08219 [hep-ex]} \BibitemShut
  {NoStop}%
\bibitem [{\citenamefont {Abe}\ \emph {et~al.}(2021)\citenamefont {Abe} \emph
  {et~al.}}]{T2K:2021xwb}%
  \BibitemOpen
  \bibfield  {author} {\bibinfo {author} {\bibfnamefont {K.}~\bibnamefont
  {Abe}} \emph {et~al.} (\bibinfo {collaboration} {T2K}),\ }\href {\doibase
  10.1103/PhysRevD.103.112008} {\bibfield  {journal} {\bibinfo  {journal}
  {Phys. Rev. D}\ }\textbf {\bibinfo {volume} {103}},\ \bibinfo {pages}
  {112008} (\bibinfo {year} {2021})},\ \Eprint
  {http://arxiv.org/abs/2101.03779} {arXiv:2101.03779 [hep-ex]} \BibitemShut
  {NoStop}%
\bibitem [{\citenamefont {Abi}\ \emph {et~al.}(2020)\citenamefont {Abi} \emph
  {et~al.}}]{DUNE:2020jqi}%
  \BibitemOpen
  \bibfield  {author} {\bibinfo {author} {\bibfnamefont {B.}~\bibnamefont
  {Abi}} \emph {et~al.} (\bibinfo {collaboration} {DUNE}),\ }\href {\doibase
  10.1140/epjc/s10052-020-08456-z} {\bibfield  {journal} {\bibinfo  {journal}
  {Eur. Phys. J. C}\ }\textbf {\bibinfo {volume} {80}},\ \bibinfo {pages} {978}
  (\bibinfo {year} {2020})},\ \Eprint {http://arxiv.org/abs/2006.16043}
  {arXiv:2006.16043 [hep-ex]} \BibitemShut {NoStop}%
\bibitem [{\citenamefont {de~Gouvea}\ and\ \citenamefont
  {Vogel}(2013)}]{deGouvea:2013zba}%
  \BibitemOpen
  \bibfield  {author} {\bibinfo {author} {\bibfnamefont {A.}~\bibnamefont
  {de~Gouvea}}\ and\ \bibinfo {author} {\bibfnamefont {P.}~\bibnamefont
  {Vogel}},\ }\href {\doibase 10.1016/j.ppnp.2013.03.006} {\bibfield  {journal}
  {\bibinfo  {journal} {Prog. Part. Nucl. Phys.}\ }\textbf {\bibinfo {volume}
  {71}},\ \bibinfo {pages} {75} (\bibinfo {year} {2013})},\ \Eprint
  {http://arxiv.org/abs/1303.4097} {arXiv:1303.4097 [hep-ph]} \BibitemShut
  {NoStop}%
\bibitem [{\citenamefont {Bernstein}\ and\ \citenamefont
  {Cooper}(2013)}]{Bernstein:2013hba}%
  \BibitemOpen
  \bibfield  {author} {\bibinfo {author} {\bibfnamefont {R.~H.}\ \bibnamefont
  {Bernstein}}\ and\ \bibinfo {author} {\bibfnamefont {P.~S.}\ \bibnamefont
  {Cooper}},\ }\href {\doibase 10.1016/j.physrep.2013.07.002} {\bibfield
  {journal} {\bibinfo  {journal} {Phys. Rept.}\ }\textbf {\bibinfo {volume}
  {532}},\ \bibinfo {pages} {27} (\bibinfo {year} {2013})},\ \Eprint
  {http://arxiv.org/abs/1307.5787} {arXiv:1307.5787 [hep-ex]} \BibitemShut
  {NoStop}%
\bibitem [{\citenamefont {Lee}(2018)}]{Lee:2018wcx}%
  \BibitemOpen
  \bibfield  {author} {\bibinfo {author} {\bibfnamefont {M.}~\bibnamefont
  {Lee}},\ }\href {\doibase 10.3389/fphy.2018.00133} {\bibfield  {journal}
  {\bibinfo  {journal} {Front. in Phys.}\ }\textbf {\bibinfo {volume} {6}}
  (\bibinfo {year} {2018}),\ 10.3389/fphy.2018.00133}\BibitemShut {NoStop}%
\bibitem [{\citenamefont {Bernstein}(2019)}]{Bernstein:2019fyh}%
  \BibitemOpen
  \bibfield  {author} {\bibinfo {author} {\bibfnamefont {R.~H.}\ \bibnamefont
  {Bernstein}} (\bibinfo {collaboration} {Mu2e}),\ }\href {\doibase
  10.3389/fphy.2019.00001} {\bibfield  {journal} {\bibinfo  {journal} {Front.
  in Phys.}\ }\textbf {\bibinfo {volume} {7}},\ \bibinfo {pages} {1} (\bibinfo
  {year} {2019})},\ \Eprint {http://arxiv.org/abs/1901.11099} {arXiv:1901.11099
  [physics.ins-det]} \BibitemShut {NoStop}%
\bibitem [{\citenamefont {Bopp}\ \emph {et~al.}(1986)\citenamefont {Bopp},
  \citenamefont {Dubbers}, \citenamefont {Hornig}, \citenamefont {Klemt},
  \citenamefont {Last}, \citenamefont {Schutze}, \citenamefont {Freedman},\
  and\ \citenamefont {Scharpf}}]{Bopp:1986rt}%
  \BibitemOpen
  \bibfield  {author} {\bibinfo {author} {\bibfnamefont {P.}~\bibnamefont
  {Bopp}}, \bibinfo {author} {\bibfnamefont {D.}~\bibnamefont {Dubbers}},
  \bibinfo {author} {\bibfnamefont {L.}~\bibnamefont {Hornig}}, \bibinfo
  {author} {\bibfnamefont {E.}~\bibnamefont {Klemt}}, \bibinfo {author}
  {\bibfnamefont {J.}~\bibnamefont {Last}}, \bibinfo {author} {\bibfnamefont
  {H.}~\bibnamefont {Schutze}}, \bibinfo {author} {\bibfnamefont {S.~J.}\
  \bibnamefont {Freedman}}, \ and\ \bibinfo {author} {\bibfnamefont
  {O.}~\bibnamefont {Scharpf}},\ }\href {\doibase 10.1103/PhysRevLett.56.919}
  {\bibfield  {journal} {\bibinfo  {journal} {Phys. Rev. Lett.}\ }\textbf
  {\bibinfo {volume} {56}},\ \bibinfo {pages} {919} (\bibinfo {year} {1986})},\
  \bibinfo {note} {[Erratum: Phys.Rev.Lett. 57, 1192 (1986)]}\BibitemShut
  {NoStop}%
\bibitem [{\citenamefont {Ando}\ \emph {et~al.}(2004)\citenamefont {Ando},
  \citenamefont {Fearing}, \citenamefont {Gudkov}, \citenamefont {Kubodera},
  \citenamefont {Myhrer}, \citenamefont {Nakamura},\ and\ \citenamefont
  {Sato}}]{Ando:2004rk}%
  \BibitemOpen
  \bibfield  {author} {\bibinfo {author} {\bibfnamefont {S.}~\bibnamefont
  {Ando}}, \bibinfo {author} {\bibfnamefont {H.~W.}\ \bibnamefont {Fearing}},
  \bibinfo {author} {\bibfnamefont {V.~P.}\ \bibnamefont {Gudkov}}, \bibinfo
  {author} {\bibfnamefont {K.}~\bibnamefont {Kubodera}}, \bibinfo {author}
  {\bibfnamefont {F.}~\bibnamefont {Myhrer}}, \bibinfo {author} {\bibfnamefont
  {S.}~\bibnamefont {Nakamura}}, \ and\ \bibinfo {author} {\bibfnamefont
  {T.}~\bibnamefont {Sato}},\ }\href {\doibase 10.1016/j.physletb.2004.06.037}
  {\bibfield  {journal} {\bibinfo  {journal} {Phys. Lett. B}\ }\textbf
  {\bibinfo {volume} {595}},\ \bibinfo {pages} {250} (\bibinfo {year}
  {2004})},\ \Eprint {http://arxiv.org/abs/nucl-th/0402100}
  {arXiv:nucl-th/0402100} \BibitemShut {NoStop}%
\bibitem [{\citenamefont {Darius}\ \emph {et~al.}(2017)\citenamefont {Darius}
  \emph {et~al.}}]{Darius:2017arh}%
  \BibitemOpen
  \bibfield  {author} {\bibinfo {author} {\bibfnamefont {G.}~\bibnamefont
  {Darius}} \emph {et~al.},\ }\href {\doibase 10.1103/PhysRevLett.119.042502}
  {\bibfield  {journal} {\bibinfo  {journal} {Phys. Rev. Lett.}\ }\textbf
  {\bibinfo {volume} {119}},\ \bibinfo {pages} {042502} (\bibinfo {year}
  {2017})}\BibitemShut {NoStop}%
\bibitem [{\citenamefont {Seng}\ \emph {et~al.}(2018)\citenamefont {Seng},
  \citenamefont {Gorchtein}, \citenamefont {Patel},\ and\ \citenamefont
  {Ramsey-Musolf}}]{Seng:2018yzq}%
  \BibitemOpen
  \bibfield  {author} {\bibinfo {author} {\bibfnamefont {C.-Y.}\ \bibnamefont
  {Seng}}, \bibinfo {author} {\bibfnamefont {M.}~\bibnamefont {Gorchtein}},
  \bibinfo {author} {\bibfnamefont {H.~H.}\ \bibnamefont {Patel}}, \ and\
  \bibinfo {author} {\bibfnamefont {M.~J.}\ \bibnamefont {Ramsey-Musolf}},\
  }\href {\doibase 10.1103/PhysRevLett.121.241804} {\bibfield  {journal}
  {\bibinfo  {journal} {Phys. Rev. Lett.}\ }\textbf {\bibinfo {volume} {121}},\
  \bibinfo {pages} {241804} (\bibinfo {year} {2018})},\ \Eprint
  {http://arxiv.org/abs/1807.10197} {arXiv:1807.10197 [hep-ph]} \BibitemShut
  {NoStop}%
\bibitem [{\citenamefont {Seng}\ \emph {et~al.}(2019)\citenamefont {Seng},
  \citenamefont {Gorchtein},\ and\ \citenamefont
  {Ramsey-Musolf}}]{Seng:2018qru}%
  \BibitemOpen
  \bibfield  {author} {\bibinfo {author} {\bibfnamefont {C.~Y.}\ \bibnamefont
  {Seng}}, \bibinfo {author} {\bibfnamefont {M.}~\bibnamefont {Gorchtein}}, \
  and\ \bibinfo {author} {\bibfnamefont {M.~J.}\ \bibnamefont
  {Ramsey-Musolf}},\ }\href {\doibase 10.1103/PhysRevD.100.013001} {\bibfield
  {journal} {\bibinfo  {journal} {Phys. Rev. D}\ }\textbf {\bibinfo {volume}
  {100}},\ \bibinfo {pages} {013001} (\bibinfo {year} {2019})},\ \Eprint
  {http://arxiv.org/abs/1812.03352} {arXiv:1812.03352 [nucl-th]} \BibitemShut
  {NoStop}%
\bibitem [{\citenamefont {Fry}\ \emph {et~al.}(2019)\citenamefont {Fry} \emph
  {et~al.}}]{Fry:2018kvq}%
  \BibitemOpen
  \bibfield  {author} {\bibinfo {author} {\bibfnamefont {J.}~\bibnamefont
  {Fry}} \emph {et~al.},\ }\href {\doibase 10.1051/epjconf/201921904002}
  {\bibfield  {journal} {\bibinfo  {journal} {EPJ Web Conf.}\ }\textbf
  {\bibinfo {volume} {219}},\ \bibinfo {pages} {04002} (\bibinfo {year}
  {2019})},\ \Eprint {http://arxiv.org/abs/1811.10047} {arXiv:1811.10047
  [nucl-ex]} \BibitemShut {NoStop}%
\bibitem [{\citenamefont {Czarnecki}\ \emph {et~al.}(2019)\citenamefont
  {Czarnecki}, \citenamefont {Marciano},\ and\ \citenamefont
  {Sirlin}}]{Czarnecki:2019mwq}%
  \BibitemOpen
  \bibfield  {author} {\bibinfo {author} {\bibfnamefont {A.}~\bibnamefont
  {Czarnecki}}, \bibinfo {author} {\bibfnamefont {W.~J.}\ \bibnamefont
  {Marciano}}, \ and\ \bibinfo {author} {\bibfnamefont {A.}~\bibnamefont
  {Sirlin}},\ }\href {\doibase 10.1103/PhysRevD.100.073008} {\bibfield
  {journal} {\bibinfo  {journal} {Phys. Rev. D}\ }\textbf {\bibinfo {volume}
  {100}},\ \bibinfo {pages} {073008} (\bibinfo {year} {2019})},\ \Eprint
  {http://arxiv.org/abs/1907.06737} {arXiv:1907.06737 [hep-ph]} \BibitemShut
  {NoStop}%
\bibitem [{\citenamefont {Hayen}(2021)}]{Hayen:2020cxh}%
  \BibitemOpen
  \bibfield  {author} {\bibinfo {author} {\bibfnamefont {L.}~\bibnamefont
  {Hayen}},\ }\href {\doibase 10.1103/PhysRevD.103.113001} {\bibfield
  {journal} {\bibinfo  {journal} {Phys. Rev. D}\ }\textbf {\bibinfo {volume}
  {103}},\ \bibinfo {pages} {113001} (\bibinfo {year} {2021})},\ \Eprint
  {http://arxiv.org/abs/2010.07262} {arXiv:2010.07262 [hep-ph]} \BibitemShut
  {NoStop}%
\bibitem [{\citenamefont {Seng}\ \emph {et~al.}(2020)\citenamefont {Seng},
  \citenamefont {Feng}, \citenamefont {Gorchtein},\ and\ \citenamefont
  {Jin}}]{Seng:2020wjq}%
  \BibitemOpen
  \bibfield  {author} {\bibinfo {author} {\bibfnamefont {C.-Y.}\ \bibnamefont
  {Seng}}, \bibinfo {author} {\bibfnamefont {X.}~\bibnamefont {Feng}}, \bibinfo
  {author} {\bibfnamefont {M.}~\bibnamefont {Gorchtein}}, \ and\ \bibinfo
  {author} {\bibfnamefont {L.-C.}\ \bibnamefont {Jin}},\ }\href {\doibase
  10.1103/PhysRevD.101.111301} {\bibfield  {journal} {\bibinfo  {journal}
  {Phys. Rev. D}\ }\textbf {\bibinfo {volume} {101}},\ \bibinfo {pages}
  {111301} (\bibinfo {year} {2020})},\ \Eprint
  {http://arxiv.org/abs/2003.11264} {arXiv:2003.11264 [hep-ph]} \BibitemShut
  {NoStop}%
\bibitem [{\citenamefont {Gorchtein}\ and\ \citenamefont
  {Seng}(2021)}]{Gorchtein:2021fce}%
  \BibitemOpen
  \bibfield  {author} {\bibinfo {author} {\bibfnamefont {M.}~\bibnamefont
  {Gorchtein}}\ and\ \bibinfo {author} {\bibfnamefont {C.-Y.}\ \bibnamefont
  {Seng}},\ }\href {\doibase 10.1007/JHEP10(2021)053} {\bibfield  {journal}
  {\bibinfo  {journal} {JHEP}\ }\textbf {\bibinfo {volume} {10}},\ \bibinfo
  {pages} {053} (\bibinfo {year} {2021})},\ \Eprint
  {http://arxiv.org/abs/2106.09185} {arXiv:2106.09185 [hep-ph]} \BibitemShut
  {NoStop}%
\bibitem [{\citenamefont {Hardy}\ and\ \citenamefont
  {Towner}(2020)}]{Hardy:2020qwl}%
  \BibitemOpen
  \bibfield  {author} {\bibinfo {author} {\bibfnamefont {J.~C.}\ \bibnamefont
  {Hardy}}\ and\ \bibinfo {author} {\bibfnamefont {I.~S.}\ \bibnamefont
  {Towner}},\ }\href {\doibase 10.1103/PhysRevC.102.045501} {\bibfield
  {journal} {\bibinfo  {journal} {Phys. Rev. C}\ }\textbf {\bibinfo {volume}
  {102}},\ \bibinfo {pages} {045501} (\bibinfo {year} {2020})}\BibitemShut
  {NoStop}%
\bibitem [{\citenamefont {Gonzalez}\ \emph {et~al.}(2021)\citenamefont
  {Gonzalez} \emph {et~al.}}]{UCNt:2021pcg}%
  \BibitemOpen
  \bibfield  {author} {\bibinfo {author} {\bibfnamefont {F.~M.}\ \bibnamefont
  {Gonzalez}} \emph {et~al.} (\bibinfo {collaboration}
  {UCN\ensuremath{\tau}}),\ }\href {\doibase 10.1103/PhysRevLett.127.162501}
  {\bibfield  {journal} {\bibinfo  {journal} {Phys. Rev. Lett.}\ }\textbf
  {\bibinfo {volume} {127}},\ \bibinfo {pages} {162501} (\bibinfo {year}
  {2021})},\ \Eprint {http://arxiv.org/abs/2106.10375} {arXiv:2106.10375
  [nucl-ex]} \BibitemShut {NoStop}%
\bibitem [{\citenamefont {Shiells}\ \emph {et~al.}(2021)\citenamefont
  {Shiells}, \citenamefont {Blunden},\ and\ \citenamefont
  {Melnitchouk}}]{Shiells:2020fqp}%
  \BibitemOpen
  \bibfield  {author} {\bibinfo {author} {\bibfnamefont {K.}~\bibnamefont
  {Shiells}}, \bibinfo {author} {\bibfnamefont {P.~G.}\ \bibnamefont
  {Blunden}}, \ and\ \bibinfo {author} {\bibfnamefont {W.}~\bibnamefont
  {Melnitchouk}},\ }\href {\doibase 10.1103/PhysRevD.104.033003} {\bibfield
  {journal} {\bibinfo  {journal} {Phys. Rev. D}\ }\textbf {\bibinfo {volume}
  {104}},\ \bibinfo {pages} {033003} (\bibinfo {year} {2021})},\ \Eprint
  {http://arxiv.org/abs/2012.01580} {arXiv:2012.01580 [hep-ph]} \BibitemShut
  {NoStop}%
\bibitem [{\citenamefont {Cirigliano}\ \emph {et~al.}(2013)\citenamefont
  {Cirigliano}, \citenamefont {Gardner},\ and\ \citenamefont
  {Holstein}}]{Cirigliano:2013xha}%
  \BibitemOpen
  \bibfield  {author} {\bibinfo {author} {\bibfnamefont {V.}~\bibnamefont
  {Cirigliano}}, \bibinfo {author} {\bibfnamefont {S.}~\bibnamefont {Gardner}},
  \ and\ \bibinfo {author} {\bibfnamefont {B.}~\bibnamefont {Holstein}},\
  }\href {\doibase 10.1016/j.ppnp.2013.03.005} {\bibfield  {journal} {\bibinfo
  {journal} {Prog. Part. Nucl. Phys.}\ }\textbf {\bibinfo {volume} {71}},\
  \bibinfo {pages} {93} (\bibinfo {year} {2013})},\ \Eprint
  {http://arxiv.org/abs/1303.6953} {arXiv:1303.6953 [hep-ph]} \BibitemShut
  {NoStop}%
\bibitem [{\citenamefont {Glick-Magid}\ \emph {et~al.}(2017)\citenamefont
  {Glick-Magid}, \citenamefont {Mishnayot}, \citenamefont {Mukul},
  \citenamefont {Hass}, \citenamefont {Vaintraub}, \citenamefont {Ron},\ and\
  \citenamefont {Gazit}}]{Glick-Magid:2016rsv}%
  \BibitemOpen
  \bibfield  {author} {\bibinfo {author} {\bibfnamefont {A.}~\bibnamefont
  {Glick-Magid}}, \bibinfo {author} {\bibfnamefont {Y.}~\bibnamefont
  {Mishnayot}}, \bibinfo {author} {\bibfnamefont {I.}~\bibnamefont {Mukul}},
  \bibinfo {author} {\bibfnamefont {M.}~\bibnamefont {Hass}}, \bibinfo {author}
  {\bibfnamefont {S.}~\bibnamefont {Vaintraub}}, \bibinfo {author}
  {\bibfnamefont {G.}~\bibnamefont {Ron}}, \ and\ \bibinfo {author}
  {\bibfnamefont {D.}~\bibnamefont {Gazit}},\ }\href {\doibase
  10.1016/j.physletb.2017.02.023} {\bibfield  {journal} {\bibinfo  {journal}
  {Phys. Lett. B}\ }\textbf {\bibinfo {volume} {767}},\ \bibinfo {pages} {285}
  (\bibinfo {year} {2017})},\ \Eprint {http://arxiv.org/abs/1609.03268}
  {arXiv:1609.03268 [nucl-ex]} \BibitemShut {NoStop}%
\bibitem [{\citenamefont {Gonz\'alez-Alonso}\ \emph {et~al.}(2019)\citenamefont
  {Gonz\'alez-Alonso}, \citenamefont {Naviliat-Cuncic},\ and\ \citenamefont
  {Severijns}}]{Gonzalez-Alonso:2018omy}%
  \BibitemOpen
  \bibfield  {author} {\bibinfo {author} {\bibfnamefont {M.}~\bibnamefont
  {Gonz\'alez-Alonso}}, \bibinfo {author} {\bibfnamefont {O.}~\bibnamefont
  {Naviliat-Cuncic}}, \ and\ \bibinfo {author} {\bibfnamefont {N.}~\bibnamefont
  {Severijns}},\ }\href {\doibase 10.1016/j.ppnp.2018.08.002} {\bibfield
  {journal} {\bibinfo  {journal} {Prog. Part. Nucl. Phys.}\ }\textbf {\bibinfo
  {volume} {104}},\ \bibinfo {pages} {165} (\bibinfo {year} {2019})},\ \Eprint
  {http://arxiv.org/abs/1803.08732} {arXiv:1803.08732 [hep-ph]} \BibitemShut
  {NoStop}%
\bibitem [{\citenamefont {Glick-Magid}\ \emph {et~al.}(2022)\citenamefont
  {Glick-Magid}, \citenamefont {Forss\'en}, \citenamefont {Gazda},
  \citenamefont {Gazit}, \citenamefont {Gysbers},\ and\ \citenamefont
  {Navr\'atil}}]{Glick-Magid:2021uwb}%
  \BibitemOpen
  \bibfield  {author} {\bibinfo {author} {\bibfnamefont {A.}~\bibnamefont
  {Glick-Magid}}, \bibinfo {author} {\bibfnamefont {C.}~\bibnamefont
  {Forss\'en}}, \bibinfo {author} {\bibfnamefont {D.}~\bibnamefont {Gazda}},
  \bibinfo {author} {\bibfnamefont {D.}~\bibnamefont {Gazit}}, \bibinfo
  {author} {\bibfnamefont {P.}~\bibnamefont {Gysbers}}, \ and\ \bibinfo
  {author} {\bibfnamefont {P.}~\bibnamefont {Navr\'atil}},\ }\href {\doibase
  10.1016/j.physletb.2022.137259} {\bibfield  {journal} {\bibinfo  {journal}
  {Phys. Lett. B}\ }\textbf {\bibinfo {volume} {832}},\ \bibinfo {pages}
  {137259} (\bibinfo {year} {2022})},\ \Eprint
  {http://arxiv.org/abs/2107.10212} {arXiv:2107.10212 [nucl-th]} \BibitemShut
  {NoStop}%
\bibitem [{\citenamefont {Falkowski}\ \emph {et~al.}(2021)\citenamefont
  {Falkowski}, \citenamefont {Gonz\'alez-Alonso}, \citenamefont {Palavri\'c},\
  and\ \citenamefont {Rodr\'\i{}guez-S\'anchez}}]{Falkowski:2021vdg}%
  \BibitemOpen
  \bibfield  {author} {\bibinfo {author} {\bibfnamefont {A.}~\bibnamefont
  {Falkowski}}, \bibinfo {author} {\bibfnamefont {M.}~\bibnamefont
  {Gonz\'alez-Alonso}}, \bibinfo {author} {\bibfnamefont {A.}~\bibnamefont
  {Palavri\'c}}, \ and\ \bibinfo {author} {\bibfnamefont {A.}~\bibnamefont
  {Rodr\'\i{}guez-S\'anchez}},\ }\href@noop {} {\  (\bibinfo {year} {2021})},\
  \Eprint {http://arxiv.org/abs/2112.07688} {arXiv:2112.07688 [hep-ph]}
  \BibitemShut {NoStop}%
\bibitem [{\citenamefont {Brodeur}\ \emph {et~al.}(2023)\citenamefont {Brodeur}
  \emph {et~al.}}]{Brodeur:2023eul}%
  \BibitemOpen
  \bibfield  {author} {\bibinfo {author} {\bibfnamefont {M.}~\bibnamefont
  {Brodeur}} \emph {et~al.}\ }(\bibinfo {year} {2023})\ \Eprint
  {http://arxiv.org/abs/2301.03975} {arXiv:2301.03975 [nucl-ex]} \BibitemShut
  {NoStop}%
\bibitem [{\citenamefont {Crivellin}\ \emph {et~al.}(2020)\citenamefont
  {Crivellin}, \citenamefont {Kirk}, \citenamefont {Manzari},\ and\
  \citenamefont {Montull}}]{Crivellin:2020ebi}%
  \BibitemOpen
  \bibfield  {author} {\bibinfo {author} {\bibfnamefont {A.}~\bibnamefont
  {Crivellin}}, \bibinfo {author} {\bibfnamefont {F.}~\bibnamefont {Kirk}},
  \bibinfo {author} {\bibfnamefont {C.~A.}\ \bibnamefont {Manzari}}, \ and\
  \bibinfo {author} {\bibfnamefont {M.}~\bibnamefont {Montull}},\ }\href
  {\doibase 10.1007/JHEP12(2020)166} {\bibfield  {journal} {\bibinfo  {journal}
  {JHEP}\ }\textbf {\bibinfo {volume} {12}},\ \bibinfo {pages} {166} (\bibinfo
  {year} {2020})},\ \Eprint {http://arxiv.org/abs/2008.01113} {arXiv:2008.01113
  [hep-ph]} \BibitemShut {NoStop}%
\bibitem [{\citenamefont {Coutinho}\ \emph {et~al.}(2020)\citenamefont
  {Coutinho}, \citenamefont {Crivellin},\ and\ \citenamefont
  {Manzari}}]{Coutinho:2019aiy}%
  \BibitemOpen
  \bibfield  {author} {\bibinfo {author} {\bibfnamefont {A.~M.}\ \bibnamefont
  {Coutinho}}, \bibinfo {author} {\bibfnamefont {A.}~\bibnamefont {Crivellin}},
  \ and\ \bibinfo {author} {\bibfnamefont {C.~A.}\ \bibnamefont {Manzari}},\
  }\href {\doibase 10.1103/PhysRevLett.125.071802} {\bibfield  {journal}
  {\bibinfo  {journal} {Phys. Rev. Lett.}\ }\textbf {\bibinfo {volume} {125}},\
  \bibinfo {pages} {071802} (\bibinfo {year} {2020})},\ \Eprint
  {http://arxiv.org/abs/1912.08823} {arXiv:1912.08823 [hep-ph]} \BibitemShut
  {NoStop}%
\bibitem [{\citenamefont {Crivellin}\ \emph {et~al.}(2021)\citenamefont
  {Crivellin}, \citenamefont {Hoferichter},\ and\ \citenamefont
  {Manzari}}]{Crivellin:2021njn}%
  \BibitemOpen
  \bibfield  {author} {\bibinfo {author} {\bibfnamefont {A.}~\bibnamefont
  {Crivellin}}, \bibinfo {author} {\bibfnamefont {M.}~\bibnamefont
  {Hoferichter}}, \ and\ \bibinfo {author} {\bibfnamefont {C.~A.}\ \bibnamefont
  {Manzari}},\ }\href {\doibase 10.1103/PhysRevLett.127.071801} {\bibfield
  {journal} {\bibinfo  {journal} {Phys. Rev. Lett.}\ }\textbf {\bibinfo
  {volume} {127}},\ \bibinfo {pages} {071801} (\bibinfo {year} {2021})},\
  \Eprint {http://arxiv.org/abs/2102.02825} {arXiv:2102.02825 [hep-ph]}
  \BibitemShut {NoStop}%
\bibitem [{\citenamefont {Crivellin}\ and\ \citenamefont
  {Hoferichter}(2020)}]{Crivellin:2020lzu}%
  \BibitemOpen
  \bibfield  {author} {\bibinfo {author} {\bibfnamefont {A.}~\bibnamefont
  {Crivellin}}\ and\ \bibinfo {author} {\bibfnamefont {M.}~\bibnamefont
  {Hoferichter}},\ }\href {\doibase 10.1103/PhysRevLett.125.111801} {\bibfield
  {journal} {\bibinfo  {journal} {Phys. Rev. Lett.}\ }\textbf {\bibinfo
  {volume} {125}},\ \bibinfo {pages} {111801} (\bibinfo {year} {2020})},\
  \Eprint {http://arxiv.org/abs/2002.07184} {arXiv:2002.07184 [hep-ph]}
  \BibitemShut {NoStop}%
\bibitem [{\citenamefont {Cirigliano}\ \emph
  {et~al.}(2023{\natexlab{a}})\citenamefont {Cirigliano}, \citenamefont
  {Crivellin}, \citenamefont {Hoferichter},\ and\ \citenamefont
  {Moulson}}]{Cirigliano:2022yyo}%
  \BibitemOpen
  \bibfield  {author} {\bibinfo {author} {\bibfnamefont {V.}~\bibnamefont
  {Cirigliano}}, \bibinfo {author} {\bibfnamefont {A.}~\bibnamefont
  {Crivellin}}, \bibinfo {author} {\bibfnamefont {M.}~\bibnamefont
  {Hoferichter}}, \ and\ \bibinfo {author} {\bibfnamefont {M.}~\bibnamefont
  {Moulson}},\ }\href {\doibase 10.1016/j.physletb.2023.137748} {\bibfield
  {journal} {\bibinfo  {journal} {Phys. Lett. B}\ }\textbf {\bibinfo {volume}
  {838}},\ \bibinfo {pages} {137748} (\bibinfo {year} {2023}{\natexlab{a}})},\
  \Eprint {http://arxiv.org/abs/2208.11707} {arXiv:2208.11707 [hep-ph]}
  \BibitemShut {NoStop}%
\bibitem [{\citenamefont {Sirlin}(1967)}]{Sirlin:1967zza}%
  \BibitemOpen
  \bibfield  {author} {\bibinfo {author} {\bibfnamefont {A.}~\bibnamefont
  {Sirlin}},\ }\href {\doibase 10.1103/PhysRev.164.1767} {\bibfield  {journal}
  {\bibinfo  {journal} {Phys. Rev.}\ }\textbf {\bibinfo {volume} {164}},\
  \bibinfo {pages} {1767} (\bibinfo {year} {1967})}\BibitemShut {NoStop}%
\bibitem [{\citenamefont {Jaus}(1972)}]{Jaus:1972hua}%
  \BibitemOpen
  \bibfield  {author} {\bibinfo {author} {\bibfnamefont {W.}~\bibnamefont
  {Jaus}},\ }\href {\doibase 10.1016/0370-2693(72)90610-7} {\bibfield
  {journal} {\bibinfo  {journal} {Phys. Lett. B}\ }\textbf {\bibinfo {volume}
  {40}},\ \bibinfo {pages} {616} (\bibinfo {year} {1972})}\BibitemShut
  {NoStop}%
\bibitem [{\citenamefont {Wilkinson}(1982)}]{Wilkinson:1982hu}%
  \BibitemOpen
  \bibfield  {author} {\bibinfo {author} {\bibfnamefont {D.~H.}\ \bibnamefont
  {Wilkinson}},\ }\href {\doibase 10.1016/0375-9474(82)90051-3} {\bibfield
  {journal} {\bibinfo  {journal} {Nucl. Phys. A}\ }\textbf {\bibinfo {volume}
  {377}},\ \bibinfo {pages} {474} (\bibinfo {year} {1982})}\BibitemShut
  {NoStop}%
\bibitem [{\citenamefont {Sirlin}\ and\ \citenamefont
  {Zucchini}(1986)}]{Sirlin:1986cc}%
  \BibitemOpen
  \bibfield  {author} {\bibinfo {author} {\bibfnamefont {A.}~\bibnamefont
  {Sirlin}}\ and\ \bibinfo {author} {\bibfnamefont {R.}~\bibnamefont
  {Zucchini}},\ }\href {\doibase 10.1103/PhysRevLett.57.1994} {\bibfield
  {journal} {\bibinfo  {journal} {Phys. Rev. Lett.}\ }\textbf {\bibinfo
  {volume} {57}},\ \bibinfo {pages} {1994} (\bibinfo {year}
  {1986})}\BibitemShut {NoStop}%
\bibitem [{\citenamefont {Jaus}\ and\ \citenamefont
  {Rasche}(1987)}]{Jaus:1986te}%
  \BibitemOpen
  \bibfield  {author} {\bibinfo {author} {\bibfnamefont {W.}~\bibnamefont
  {Jaus}}\ and\ \bibinfo {author} {\bibfnamefont {G.}~\bibnamefont {Rasche}},\
  }\href {\doibase 10.1103/PhysRevD.35.3420} {\bibfield  {journal} {\bibinfo
  {journal} {Phys. Rev. D}\ }\textbf {\bibinfo {volume} {35}},\ \bibinfo
  {pages} {3420} (\bibinfo {year} {1987})}\BibitemShut {NoStop}%
\bibitem [{\citenamefont {Cirigliano}\ \emph {et~al.}(2022)\citenamefont
  {Cirigliano}, \citenamefont {de~Vries}, \citenamefont {Hayen}, \citenamefont
  {Mereghetti},\ and\ \citenamefont {Walker-Loud}}]{Cirigliano:2022hob}%
  \BibitemOpen
  \bibfield  {author} {\bibinfo {author} {\bibfnamefont {V.}~\bibnamefont
  {Cirigliano}}, \bibinfo {author} {\bibfnamefont {J.}~\bibnamefont
  {de~Vries}}, \bibinfo {author} {\bibfnamefont {L.}~\bibnamefont {Hayen}},
  \bibinfo {author} {\bibfnamefont {E.}~\bibnamefont {Mereghetti}}, \ and\
  \bibinfo {author} {\bibfnamefont {A.}~\bibnamefont {Walker-Loud}},\ }\href
  {\doibase 10.1103/PhysRevLett.129.121801} {\bibfield  {journal} {\bibinfo
  {journal} {Phys. Rev. Lett.}\ }\textbf {\bibinfo {volume} {129}},\ \bibinfo
  {pages} {121801} (\bibinfo {year} {2022})},\ \Eprint
  {http://arxiv.org/abs/2202.10439} {arXiv:2202.10439 [nucl-th]} \BibitemShut
  {NoStop}%
\bibitem [{\citenamefont {Gorchtein}\ and\ \citenamefont
  {Seng}(2023)}]{Gorchtein:2023naa}%
  \BibitemOpen
  \bibfield  {author} {\bibinfo {author} {\bibfnamefont {M.}~\bibnamefont
  {Gorchtein}}\ and\ \bibinfo {author} {\bibfnamefont {C.~Y.}\ \bibnamefont
  {Seng}},\ }\href@noop {} {\  (\bibinfo {year} {2023})},\ \Eprint
  {http://arxiv.org/abs/2311.00044} {arXiv:2311.00044 [nucl-th]} \BibitemShut
  {NoStop}%
\bibitem [{\citenamefont {Borah}\ \emph {et~al.}(2024)\citenamefont {Borah},
  \citenamefont {Hill},\ and\ \citenamefont {Plestid}}]{z2a3anom}%
  \BibitemOpen
  \bibfield  {author} {\bibinfo {author} {\bibfnamefont {K.}~\bibnamefont
  {Borah}}, \bibinfo {author} {\bibfnamefont {R.~J.}\ \bibnamefont {Hill}}, \
  and\ \bibinfo {author} {\bibfnamefont {R.}~\bibnamefont {Plestid}},\
  }\href@noop {} {\  (\bibinfo {year} {2024})},\ \Eprint
  {http://arxiv.org/abs/2402.13307} {arXiv:2402.13307 [hep-ph]} \BibitemShut
  {NoStop}%
\bibitem [{\citenamefont {Hill}\ and\ \citenamefont
  {Plestid}(2023)}]{Hill:2023bfh}%
  \BibitemOpen
  \bibfield  {author} {\bibinfo {author} {\bibfnamefont {R.~J.}\ \bibnamefont
  {Hill}}\ and\ \bibinfo {author} {\bibfnamefont {R.}~\bibnamefont {Plestid}},\
  }\href@noop {} {\  (\bibinfo {year} {2023})},\ \Eprint
  {http://arxiv.org/abs/2309.15929} {arXiv:2309.15929 [hep-ph]} \BibitemShut
  {NoStop}%
\bibitem [{\citenamefont {{Vander Griend}}\ \emph {et~al.}(2023)\citenamefont
  {{Vander Griend}}, \citenamefont {Hill},\ and\ \citenamefont
  {Plestid}}]{largepi}%
  \BibitemOpen
  \bibfield  {author} {\bibinfo {author} {\bibfnamefont {P.}~\bibnamefont
  {{Vander Griend}}}, \bibinfo {author} {\bibfnamefont {R.~J.}\ \bibnamefont
  {Hill}}, \ and\ \bibinfo {author} {\bibfnamefont {R.}~\bibnamefont
  {Plestid}},\ }\href@noop {} {\enquote {\bibinfo {title} {{Resummation of
  $\pi$-enhancements and neutron beta decay}},}\ }\bibinfo {howpublished} {In
  preparation} (\bibinfo {year} {2023})\BibitemShut {NoStop}%
\bibitem [{\citenamefont {Fermi}(1934)}]{Fermi:1934hr}%
  \BibitemOpen
  \bibfield  {author} {\bibinfo {author} {\bibfnamefont {E.}~\bibnamefont
  {Fermi}},\ }\href {\doibase 10.1007/BF01351864} {\bibfield  {journal}
  {\bibinfo  {journal} {Z. Phys.}\ }\textbf {\bibinfo {volume} {88}},\ \bibinfo
  {pages} {161} (\bibinfo {year} {1934})}\BibitemShut {NoStop}%
\bibitem [{\citenamefont {Hayen}\ \emph {et~al.}(2018)\citenamefont {Hayen},
  \citenamefont {Severijns}, \citenamefont {Bodek}, \citenamefont {Rozpedzik},\
  and\ \citenamefont {Mougeot}}]{Hayen:2017pwg}%
  \BibitemOpen
  \bibfield  {author} {\bibinfo {author} {\bibfnamefont {L.}~\bibnamefont
  {Hayen}}, \bibinfo {author} {\bibfnamefont {N.}~\bibnamefont {Severijns}},
  \bibinfo {author} {\bibfnamefont {K.}~\bibnamefont {Bodek}}, \bibinfo
  {author} {\bibfnamefont {D.}~\bibnamefont {Rozpedzik}}, \ and\ \bibinfo
  {author} {\bibfnamefont {X.}~\bibnamefont {Mougeot}},\ }\href {\doibase
  10.1103/RevModPhys.90.015008} {\bibfield  {journal} {\bibinfo  {journal}
  {Rev. Mod. Phys.}\ }\textbf {\bibinfo {volume} {90}},\ \bibinfo {pages}
  {015008} (\bibinfo {year} {2018})},\ \Eprint
  {http://arxiv.org/abs/1709.07530} {arXiv:1709.07530 [nucl-th]} \BibitemShut
  {NoStop}%
\bibitem [{Note1()}]{Note1}%
  \BibitemOpen
  \bibinfo {note} {The ambiguity in defining $r$ is often resolved by including
  an explicit nuclear charge density distribution~\cite {Hayen:2017pwg}. This
  procedure does not address how to interface the Fermi function with EFT
  matching conditions, or how to properly interface with higher-order radiative
  corrections.}\BibitemShut {Stop}%
\bibitem [{\citenamefont {Sirlin}(1987)}]{Sirlin:1986hpu}%
  \BibitemOpen
  \bibfield  {author} {\bibinfo {author} {\bibfnamefont {A.}~\bibnamefont
  {Sirlin}},\ }\href {\doibase 10.1103/PhysRevD.35.3423} {\bibfield  {journal}
  {\bibinfo  {journal} {Phys. Rev. D}\ }\textbf {\bibinfo {volume} {35}},\
  \bibinfo {pages} {3423} (\bibinfo {year} {1987})}\BibitemShut {NoStop}%
\bibitem [{\citenamefont {Collins}\ \emph {et~al.}(1989)\citenamefont
  {Collins}, \citenamefont {Soper},\ and\ \citenamefont
  {Sterman}}]{Collins:1989gx}%
  \BibitemOpen
  \bibfield  {author} {\bibinfo {author} {\bibfnamefont {J.~C.}\ \bibnamefont
  {Collins}}, \bibinfo {author} {\bibfnamefont {D.~E.}\ \bibnamefont {Soper}},
  \ and\ \bibinfo {author} {\bibfnamefont {G.~F.}\ \bibnamefont {Sterman}},\
  }\href {\doibase 10.1142/9789814503266_0001} {\bibfield  {journal} {\bibinfo
  {journal} {Adv. Ser. Direct. High Energy Phys.}\ }\textbf {\bibinfo {volume}
  {5}},\ \bibinfo {pages} {1} (\bibinfo {year} {1989})},\ \Eprint
  {http://arxiv.org/abs/hep-ph/0409313} {arXiv:hep-ph/0409313} \BibitemShut
  {NoStop}%
\bibitem [{\citenamefont {Bodwin}\ \emph {et~al.}(1995)\citenamefont {Bodwin},
  \citenamefont {Braaten},\ and\ \citenamefont {Lepage}}]{Bodwin:1994jh}%
  \BibitemOpen
  \bibfield  {author} {\bibinfo {author} {\bibfnamefont {G.~T.}\ \bibnamefont
  {Bodwin}}, \bibinfo {author} {\bibfnamefont {E.}~\bibnamefont {Braaten}}, \
  and\ \bibinfo {author} {\bibfnamefont {G.~P.}\ \bibnamefont {Lepage}},\
  }\href {\doibase 10.1103/PhysRevD.55.5853} {\bibfield  {journal} {\bibinfo
  {journal} {Phys. Rev. D}\ }\textbf {\bibinfo {volume} {51}},\ \bibinfo
  {pages} {1125} (\bibinfo {year} {1995})},\ \bibinfo {note} {[Erratum:
  Phys.Rev.D 55, 5853 (1997)]},\ \Eprint {http://arxiv.org/abs/hep-ph/9407339}
  {arXiv:hep-ph/9407339} \BibitemShut {NoStop}%
\bibitem [{\citenamefont {Bauer}\ \emph {et~al.}(2002)\citenamefont {Bauer},
  \citenamefont {Fleming}, \citenamefont {Pirjol}, \citenamefont {Rothstein},\
  and\ \citenamefont {Stewart}}]{Bauer:2002nz}%
  \BibitemOpen
  \bibfield  {author} {\bibinfo {author} {\bibfnamefont {C.~W.}\ \bibnamefont
  {Bauer}}, \bibinfo {author} {\bibfnamefont {S.}~\bibnamefont {Fleming}},
  \bibinfo {author} {\bibfnamefont {D.}~\bibnamefont {Pirjol}}, \bibinfo
  {author} {\bibfnamefont {I.~Z.}\ \bibnamefont {Rothstein}}, \ and\ \bibinfo
  {author} {\bibfnamefont {I.~W.}\ \bibnamefont {Stewart}},\ }\href {\doibase
  10.1103/PhysRevD.66.014017} {\bibfield  {journal} {\bibinfo  {journal} {Phys.
  Rev. D}\ }\textbf {\bibinfo {volume} {66}},\ \bibinfo {pages} {014017}
  (\bibinfo {year} {2002})},\ \Eprint {http://arxiv.org/abs/hep-ph/0202088}
  {arXiv:hep-ph/0202088} \BibitemShut {NoStop}%
\bibitem [{Note2()}]{Note2}%
  \BibitemOpen
  \bibinfo {note} {See Supplemental Material [URL-HERE], which includes
  Ref.~\cite {Isgur:1990yhj}, for a discussion of the factorization theorem and
  how real radiation is included.}\BibitemShut {Stop}%
\bibitem [{Note3()}]{Note3}%
  \BibitemOpen
  \bibinfo {note} {The parameter $\lambda $ may be interpreted physically as
  the scale of atomic screening, which influences decay rates at sub-leading
  power~\cite {Hayen:2017pwg}.}\BibitemShut {Stop}%
\bibitem [{Note4()}]{Note4}%
  \BibitemOpen
  \bibinfo {note} {The classic Sommerfeld enhancement is obtained in a
  perturbative expansion by replacing the Dirac structures in the last line of
  (\ref {eq:DCgeneral}) with $2m$ in every propagator. This renders the problem
  UV finite \cite {Hill:2023bfh}.}\BibitemShut {Stop}%
\bibitem [{\citenamefont {Yennie}\ \emph {et~al.}(1961)\citenamefont {Yennie},
  \citenamefont {Frautschi},\ and\ \citenamefont {Suura}}]{Yennie:1961ad}%
  \BibitemOpen
  \bibfield  {author} {\bibinfo {author} {\bibfnamefont {D.}~\bibnamefont
  {Yennie}}, \bibinfo {author} {\bibfnamefont {S.~C.}\ \bibnamefont
  {Frautschi}}, \ and\ \bibinfo {author} {\bibfnamefont {H.}~\bibnamefont
  {Suura}},\ }\href {\doibase 10.1016/0003-4916(61)90151-8} {\bibfield
  {journal} {\bibinfo  {journal} {Annals Phys.}\ }\textbf {\bibinfo {volume}
  {13}},\ \bibinfo {pages} {379} (\bibinfo {year} {1961})}\BibitemShut
  {NoStop}%
\bibitem [{\citenamefont {Weinberg}(1965)}]{Weinberg:1965nx}%
  \BibitemOpen
  \bibfield  {author} {\bibinfo {author} {\bibfnamefont {S.}~\bibnamefont
  {Weinberg}},\ }\href {\doibase 10.1103/PhysRev.140.B516} {\bibfield
  {journal} {\bibinfo  {journal} {Phys. Rev.}\ }\textbf {\bibinfo {volume}
  {140}},\ \bibinfo {pages} {B516} (\bibinfo {year} {1965})}\BibitemShut
  {NoStop}%
\bibitem [{Note5()}]{Note5}%
  \BibitemOpen
  \bibinfo {note} {Covariant expressions are obtained by replacing $\beta
  \rightarrow \protect \sqrt {1-E^2/m^2}$, $\gamma _0 \rightarrow v_\mu \gamma
  ^\mu $, and $E\rightarrow v_\mu p^\mu $ where $v_\mu $ is the reference
  vector introduced in Eqs.\ (\ref {eq:superallowedL}) and (\ref
  {eq:neutronL}).}\BibitemShut {Stop}%
\bibitem [{\citenamefont {Caswell}\ and\ \citenamefont
  {Lepage}(1986)}]{Caswell:1985ui}%
  \BibitemOpen
  \bibfield  {author} {\bibinfo {author} {\bibfnamefont {W.~E.}\ \bibnamefont
  {Caswell}}\ and\ \bibinfo {author} {\bibfnamefont {G.~P.}\ \bibnamefont
  {Lepage}},\ }\href {\doibase 10.1016/0370-2693(86)91297-9} {\bibfield
  {journal} {\bibinfo  {journal} {Phys. Lett. B}\ }\textbf {\bibinfo {volume}
  {167}},\ \bibinfo {pages} {437} (\bibinfo {year} {1986})}\BibitemShut
  {NoStop}%
\bibitem [{\citenamefont {Georgi}(1990)}]{Georgi:1990um}%
  \BibitemOpen
  \bibfield  {author} {\bibinfo {author} {\bibfnamefont {H.}~\bibnamefont
  {Georgi}},\ }\href {\doibase 10.1016/0370-2693(90)91128-X} {\bibfield
  {journal} {\bibinfo  {journal} {Phys. Lett. B}\ }\textbf {\bibinfo {volume}
  {240}},\ \bibinfo {pages} {447} (\bibinfo {year} {1990})}\BibitemShut
  {NoStop}%
\bibitem [{\citenamefont {Manohar}\ and\ \citenamefont
  {Wise}(2000)}]{Manohar:2000dt}%
  \BibitemOpen
  \bibfield  {author} {\bibinfo {author} {\bibfnamefont {A.~V.}\ \bibnamefont
  {Manohar}}\ and\ \bibinfo {author} {\bibfnamefont {M.~B.}\ \bibnamefont
  {Wise}},\ }\href@noop {} {\emph {\bibinfo {title} {{Heavy quark physics}}}},\
  Vol.~\bibinfo {volume} {10}\ (\bibinfo {year} {2000})\BibitemShut {NoStop}%
\bibitem [{\citenamefont {Paz}(2015)}]{Paz:2015uga}%
  \BibitemOpen
  \bibfield  {author} {\bibinfo {author} {\bibfnamefont {G.}~\bibnamefont
  {Paz}},\ }\href {\doibase 10.1142/S021773231550128X} {\bibfield  {journal}
  {\bibinfo  {journal} {Mod. Phys. Lett. A}\ }\textbf {\bibinfo {volume}
  {30}},\ \bibinfo {pages} {1550128} (\bibinfo {year} {2015})},\ \Eprint
  {http://arxiv.org/abs/1503.07216} {arXiv:1503.07216 [hep-ph]} \BibitemShut
  {NoStop}%
\bibitem [{Note6()}]{Note6}%
  \BibitemOpen
  \bibinfo {note} {This can be shown explicitly using heavy particle Feynman
  rules in Coulomb gauge \cite {eikonal_algebra}, and confirms an old theorem
  due to Sirlin and B\'eg~\cite {Beg:1969zu}.}\BibitemShut {Stop}%
\bibitem [{Note7()}]{Note7}%
  \BibitemOpen
  \bibinfo {note} {More generally, the anomalous dimension at a given order $n$
  in $\alpha $ is a linear combination of powers $Z^i(1+Z)^i$, where $2i\le n$.
  This is a consequence of symmetries that emerge in the limit of a massless
  lepton~\cite {z2a3anom}.}\BibitemShut {Stop}%
\bibitem [{Note8()}]{Note8}%
  \BibitemOpen
  \bibinfo {note} {The relevant results are $\gamma _0^{(1)}=-3$~\cite
  {Sirlin:1967zza} at one-loop, $\gamma _1^{(2)}= -16 \zeta _2 + \protect \frac
  52 + \protect \frac {10}{3} n_e$ at two-loops~\cite {Ji:1991pr}, and $\gamma
  _2^{(3)} = -80 \zeta _4 -36 \zeta _3 + 64\zeta _2 -\protect \frac {37}{2} +
  n_e\left ( -\protect \frac {176}{3}\zeta _3 + \protect \frac {448}{9}\zeta _2
  + \protect \frac {470}{9} \right ) + \protect \frac {140}{27} n_e^2$ at
  three-loops~\cite {Chetyrkin:2003vi} with $n_e=1$ for the number of dynamical
  charged leptons. The application of $\gamma _0^{(1)}$, $\gamma _1^{(2)}$ and
  $\gamma _2^{(3)}$ to neutron beta decay has been discussed in Ref.~\cite
  {Cirigliano:2023fnz}.}\BibitemShut {Stop}%
\bibitem [{sup()}]{supp_mat_2}%
  \BibitemOpen
  \href@noop {} {}\bibinfo {note} {See Supplemental Material [URL-HERE] for
  analytic expressions of the renormalization-group improved result for the
  ratio of Wilson coefficients.}\BibitemShut {Stop}%
\bibitem [{Note9()}]{Note9}%
  \BibitemOpen
  \bibinfo {note} {Analytic results for the QED beta function are available
  through five loops~\cite {Baikov:2012zm,Herzog:2017ohr}. For our purposes we
  require $\beta _0=-(4/3) n_f$ and $\beta _1=-4 n_f$ (for $n_f=1$), {\protect
  \it cf.} Ref.~\cite {Hill:2016gdf}.}\BibitemShut {Stop}%
\bibitem [{Note10()}]{Note10}%
  \BibitemOpen
  \bibinfo {note} {A conventional choice for $\Lambda _{\protect \rm nuc.}$ is
  given in terms of the nuclear RMS charge radius, $\Lambda \to \protect \sqrt
  {6}/R$~\cite {Sirlin:1986hpu, Hardy:2004id,Hayen:2017pwg}. For numerical
  illustrations, see the Supplemental Material.}\BibitemShut {Stop}%
\bibitem [{Note11()}]{Note11}%
  \BibitemOpen
  \bibinfo {note} {This exact result replaces the ansatz of Wilkinson~\cite
  {WILKINSON1997275,Hayen:2017pwg}, with differences beginning at order
  $Z^3\alpha ^4$.}\BibitemShut {Stop}%
\bibitem [{Note12()}]{Note12}%
  \BibitemOpen
  \bibinfo {note} {The heuristic estimate~\cite {Sirlin:1986cc,Sirlin:1986hpu}
  uses results from Ref.~\cite {Jaus:1972hua,Jaus:1986te} and is defined
  relative to the contributions contained in $F(Z,E)(1+\delta _1 + \delta _2)$,
  where $F(Z,E)$ is the classical Fermi function, $\delta _1$ determines the
  order $\alpha $ correction, and $\delta _2$ determines the $Z\alpha ^2$
  correction.}\BibitemShut {Stop}%
\bibitem [{Note13()}]{Note13}%
  \BibitemOpen
  \bibinfo {note} {We include the full ${\protect \cal O}(\alpha )$ matrix
  element~\cite {Sirlin:1967zza} (translated to $\protect \overline {\protect
  \rm MS}$ in heavy particle effective theory). For this comparison we include
  only the scale-dependent logarithm, $\log (\mu /m_e)$, in the $\alpha
  (Z\alpha )$ matrix element~\cite {Sirlin:1986cc}.}\BibitemShut {Stop}%
\bibitem [{Note14()}]{Note14}%
  \BibitemOpen
  \bibinfo {note} {These matrix elements have been calculated in Ref.~\cite
  {Sirlin:1986cc}, but not in the point-like effective theory studied
  here.}\BibitemShut {Stop}%
\bibitem [{Note15()}]{Note15}%
  \BibitemOpen
  \bibinfo {note} {The two-loop anomalous dimension $\gamma _1$ and other
  subleading two-loop corrections were omitted in the leading-logarithm
  analysis of Ref.~\cite {Czarnecki:2004cw}. Ref.~\cite {Cirigliano:2023fnz}
  included $\gamma _1$ in the RG flow between nucleon and electron-mass
  scales.}\BibitemShut {Stop}%
\bibitem [{Note16()}]{Note16}%
  \BibitemOpen
  \bibinfo {note} {At tree level, less than $0.1\%$ of the total decay rate
  involves electron velocity $\beta <0.1$, and less than $10\%$ of the rate
  involves $\beta <0.5$.}\BibitemShut {Stop}%
\bibitem [{\citenamefont {Ahrens}\ \emph
  {et~al.}(2009{\natexlab{a}})\citenamefont {Ahrens}, \citenamefont {Becher},
  \citenamefont {Neubert},\ and\ \citenamefont {Yang}}]{Ahrens:2008qu}%
  \BibitemOpen
  \bibfield  {author} {\bibinfo {author} {\bibfnamefont {V.}~\bibnamefont
  {Ahrens}}, \bibinfo {author} {\bibfnamefont {T.}~\bibnamefont {Becher}},
  \bibinfo {author} {\bibfnamefont {M.}~\bibnamefont {Neubert}}, \ and\
  \bibinfo {author} {\bibfnamefont {L.~L.}\ \bibnamefont {Yang}},\ }\href
  {\doibase 10.1103/PhysRevD.79.033013} {\bibfield  {journal} {\bibinfo
  {journal} {Phys. Rev. D}\ }\textbf {\bibinfo {volume} {79}},\ \bibinfo
  {pages} {033013} (\bibinfo {year} {2009}{\natexlab{a}})},\ \Eprint
  {http://arxiv.org/abs/0808.3008} {arXiv:0808.3008 [hep-ph]} \BibitemShut
  {NoStop}%
\bibitem [{\citenamefont {Ahrens}\ \emph
  {et~al.}(2009{\natexlab{b}})\citenamefont {Ahrens}, \citenamefont {Becher},
  \citenamefont {Neubert},\ and\ \citenamefont {Yang}}]{Ahrens:2009cxz}%
  \BibitemOpen
  \bibfield  {author} {\bibinfo {author} {\bibfnamefont {V.}~\bibnamefont
  {Ahrens}}, \bibinfo {author} {\bibfnamefont {T.}~\bibnamefont {Becher}},
  \bibinfo {author} {\bibfnamefont {M.}~\bibnamefont {Neubert}}, \ and\
  \bibinfo {author} {\bibfnamefont {L.~L.}\ \bibnamefont {Yang}},\ }\href
  {\doibase 10.1140/epjc/s10052-009-1030-2} {\bibfield  {journal} {\bibinfo
  {journal} {Eur. Phys. J. C}\ }\textbf {\bibinfo {volume} {62}},\ \bibinfo
  {pages} {333} (\bibinfo {year} {2009}{\natexlab{b}})},\ \Eprint
  {http://arxiv.org/abs/0809.4283} {arXiv:0809.4283 [hep-ph]} \BibitemShut
  {NoStop}%
\bibitem [{\citenamefont {Cirigliano}\ \emph
  {et~al.}(2023{\natexlab{b}})\citenamefont {Cirigliano}, \citenamefont
  {Dekens}, \citenamefont {Mereghetti},\ and\ \citenamefont
  {Tomalak}}]{Cirigliano:2023fnz}%
  \BibitemOpen
  \bibfield  {author} {\bibinfo {author} {\bibfnamefont {V.}~\bibnamefont
  {Cirigliano}}, \bibinfo {author} {\bibfnamefont {W.}~\bibnamefont {Dekens}},
  \bibinfo {author} {\bibfnamefont {E.}~\bibnamefont {Mereghetti}}, \ and\
  \bibinfo {author} {\bibfnamefont {O.}~\bibnamefont {Tomalak}},\ }\href@noop
  {} {\  (\bibinfo {year} {2023}{\natexlab{b}})},\ \Eprint
  {http://arxiv.org/abs/2306.03138} {arXiv:2306.03138 [hep-ph]} \BibitemShut
  {NoStop}%
\bibitem [{Note17()}]{Note17}%
  \BibitemOpen
  \bibinfo {note} {Existing order $Z^2\alpha ^3$calculations~\cite
  {Jaus:1972hua} included only a subset of diagrams and are therefore
  incomplete estimates.}\BibitemShut {Stop}%
\bibitem [{\citenamefont {Wilkinson}(1997)}]{WILKINSON1997275}%
  \BibitemOpen
  \bibfield  {author} {\bibinfo {author} {\bibfnamefont {D.}~\bibnamefont
  {Wilkinson}},\ }\href {\doibase
  https://doi.org/10.1016/S0168-9002(97)00971-6} {\bibfield  {journal}
  {\bibinfo  {journal} {Nuclear Instruments and Methods in Physics Research
  Section A: Accelerators, Spectrometers, Detectors and Associated Equipment}\
  }\textbf {\bibinfo {volume} {401}},\ \bibinfo {pages} {275} (\bibinfo {year}
  {1997})}\BibitemShut {NoStop}%
\bibitem [{\citenamefont {Brown}\ \emph {et~al.}(2018)\citenamefont {Brown}
  \emph {et~al.}}]{UCNA:2017obv}%
  \BibitemOpen
  \bibfield  {author} {\bibinfo {author} {\bibfnamefont {M.~A.~P.}\
  \bibnamefont {Brown}} \emph {et~al.} (\bibinfo {collaboration} {UCNA}),\
  }\href {\doibase 10.1103/PhysRevC.97.035505} {\bibfield  {journal} {\bibinfo
  {journal} {Phys. Rev. C}\ }\textbf {\bibinfo {volume} {97}},\ \bibinfo
  {pages} {035505} (\bibinfo {year} {2018})},\ \Eprint
  {http://arxiv.org/abs/1712.00884} {arXiv:1712.00884 [nucl-ex]} \BibitemShut
  {NoStop}%
\bibitem [{\citenamefont {Shidling}\ \emph {et~al.}(2014)\citenamefont
  {Shidling} \emph {et~al.}}]{Shidling:2014ura}%
  \BibitemOpen
  \bibfield  {author} {\bibinfo {author} {\bibfnamefont {P.~D.}\ \bibnamefont
  {Shidling}} \emph {et~al.},\ }\href {\doibase 10.1103/PhysRevC.90.032501}
  {\bibfield  {journal} {\bibinfo  {journal} {Phys. Rev. C}\ }\textbf {\bibinfo
  {volume} {90}},\ \bibinfo {pages} {032501} (\bibinfo {year} {2014})},\
  \Eprint {http://arxiv.org/abs/1407.1742} {arXiv:1407.1742 [nucl-ex]}
  \BibitemShut {NoStop}%
\bibitem [{\citenamefont {Eibach}\ \emph {et~al.}(2015)\citenamefont {Eibach}
  \emph {et~al.}}]{Eibach:2015ksa}%
  \BibitemOpen
  \bibfield  {author} {\bibinfo {author} {\bibfnamefont {M.}~\bibnamefont
  {Eibach}} \emph {et~al.},\ }\href {\doibase 10.1103/PhysRevC.92.045502}
  {\bibfield  {journal} {\bibinfo  {journal} {Phys. Rev. C}\ }\textbf {\bibinfo
  {volume} {92}},\ \bibinfo {pages} {045502} (\bibinfo {year}
  {2015})}\BibitemShut {NoStop}%
\bibitem [{\citenamefont {Sternberg}\ \emph {et~al.}(2015)\citenamefont
  {Sternberg} \emph {et~al.}}]{Sternberg:2015nnr}%
  \BibitemOpen
  \bibfield  {author} {\bibinfo {author} {\bibfnamefont {M.~G.}\ \bibnamefont
  {Sternberg}} \emph {et~al.},\ }\href {\doibase
  10.1103/PhysRevLett.115.182501} {\bibfield  {journal} {\bibinfo  {journal}
  {Phys. Rev. Lett.}\ }\textbf {\bibinfo {volume} {115}},\ \bibinfo {pages}
  {182501} (\bibinfo {year} {2015})}\BibitemShut {NoStop}%
\bibitem [{\citenamefont {Gulyuz}\ \emph {et~al.}(2016)\citenamefont {Gulyuz}
  \emph {et~al.}}]{Gulyuz:2016ppg}%
  \BibitemOpen
  \bibfield  {author} {\bibinfo {author} {\bibfnamefont {K.}~\bibnamefont
  {Gulyuz}} \emph {et~al.},\ }\href {\doibase 10.1103/PhysRevLett.116.012501}
  {\bibfield  {journal} {\bibinfo  {journal} {Phys. Rev. Lett.}\ }\textbf
  {\bibinfo {volume} {116}},\ \bibinfo {pages} {012501} (\bibinfo {year}
  {2016})}\BibitemShut {NoStop}%
\bibitem [{\citenamefont {Fenker}\ \emph {et~al.}(2016)\citenamefont {Fenker}
  \emph {et~al.}}]{Fenker:2016mka}%
  \BibitemOpen
  \bibfield  {author} {\bibinfo {author} {\bibfnamefont {B.}~\bibnamefont
  {Fenker}} \emph {et~al.},\ }\href {\doibase 10.1088/1367-2630/18/7/073028}
  {\bibfield  {journal} {\bibinfo  {journal} {New J. Phys.}\ }\textbf {\bibinfo
  {volume} {18}},\ \bibinfo {pages} {073028} (\bibinfo {year} {2016})},\
  \Eprint {http://arxiv.org/abs/1602.04526} {arXiv:1602.04526
  [physics.atom-ph]} \BibitemShut {NoStop}%
\bibitem [{\citenamefont {Long}\ \emph {et~al.}(2017)\citenamefont {Long} \emph
  {et~al.}}]{Long:2017gdh}%
  \BibitemOpen
  \bibfield  {author} {\bibinfo {author} {\bibfnamefont {J.}~\bibnamefont
  {Long}} \emph {et~al.},\ }\href {\doibase 10.1103/PhysRevC.96.015502}
  {\bibfield  {journal} {\bibinfo  {journal} {Phys. Rev. C}\ }\textbf {\bibinfo
  {volume} {96}},\ \bibinfo {pages} {015502} (\bibinfo {year}
  {2017})}\BibitemShut {NoStop}%
\bibitem [{\citenamefont {Fenker}\ \emph {et~al.}(2018)\citenamefont {Fenker}
  \emph {et~al.}}]{Fenker:2017rcx}%
  \BibitemOpen
  \bibfield  {author} {\bibinfo {author} {\bibfnamefont {B.}~\bibnamefont
  {Fenker}} \emph {et~al.},\ }\href {\doibase 10.1103/PhysRevLett.120.062502}
  {\bibfield  {journal} {\bibinfo  {journal} {Phys. Rev. Lett.}\ }\textbf
  {\bibinfo {volume} {120}},\ \bibinfo {pages} {062502} (\bibinfo {year}
  {2018})},\ \Eprint {http://arxiv.org/abs/1706.00414} {arXiv:1706.00414
  [nucl-ex]} \BibitemShut {NoStop}%
\bibitem [{\citenamefont {Brodeur}\ \emph {et~al.}(2016)\citenamefont
  {Brodeur}, \citenamefont {Kelly}, \citenamefont {Long}, \citenamefont
  {Nicoloff},\ and\ \citenamefont {Schultz}}]{Brodeur:2016cci}%
  \BibitemOpen
  \bibfield  {author} {\bibinfo {author} {\bibfnamefont {M.}~\bibnamefont
  {Brodeur}}, \bibinfo {author} {\bibfnamefont {J.}~\bibnamefont {Kelly}},
  \bibinfo {author} {\bibfnamefont {J.}~\bibnamefont {Long}}, \bibinfo {author}
  {\bibfnamefont {C.}~\bibnamefont {Nicoloff}}, \ and\ \bibinfo {author}
  {\bibfnamefont {B.}~\bibnamefont {Schultz}},\ }\href {\doibase
  10.1016/j.nimb.2015.12.038} {\bibfield  {journal} {\bibinfo  {journal} {Nucl.
  Instrum. Meth. B}\ }\textbf {\bibinfo {volume} {376}},\ \bibinfo {pages}
  {281} (\bibinfo {year} {2016})}\BibitemShut {NoStop}%
\bibitem [{\citenamefont {Shidling}\ \emph {et~al.}(2018)\citenamefont
  {Shidling} \emph {et~al.}}]{Shidling:2018fvb}%
  \BibitemOpen
  \bibfield  {author} {\bibinfo {author} {\bibfnamefont {P.~D.}\ \bibnamefont
  {Shidling}} \emph {et~al.},\ }\href {\doibase 10.1103/PhysRevC.98.015502}
  {\bibfield  {journal} {\bibinfo  {journal} {Phys. Rev. C}\ }\textbf {\bibinfo
  {volume} {98}},\ \bibinfo {pages} {015502} (\bibinfo {year}
  {2018})}\BibitemShut {NoStop}%
\bibitem [{\citenamefont {Valverde}\ \emph {et~al.}(2018)\citenamefont
  {Valverde} \emph {et~al.}}]{Valverde:2018haz}%
  \BibitemOpen
  \bibfield  {author} {\bibinfo {author} {\bibfnamefont {A.~A.}\ \bibnamefont
  {Valverde}} \emph {et~al.},\ }\href {\doibase 10.1103/PhysRevC.97.035503}
  {\bibfield  {journal} {\bibinfo  {journal} {Phys. Rev. C}\ }\textbf {\bibinfo
  {volume} {97}},\ \bibinfo {pages} {035503} (\bibinfo {year}
  {2018})}\BibitemShut {NoStop}%
\bibitem [{\citenamefont {O'Malley}\ \emph {et~al.}(2020)\citenamefont
  {O'Malley} \emph {et~al.}}]{OMalley:2020vop}%
  \BibitemOpen
  \bibfield  {author} {\bibinfo {author} {\bibfnamefont {P.~D.}\ \bibnamefont
  {O'Malley}} \emph {et~al.},\ }\href {\doibase 10.1016/j.nimb.2019.04.017}
  {\bibfield  {journal} {\bibinfo  {journal} {Nucl. Instrum. Meth. B}\ }\textbf
  {\bibinfo {volume} {463}},\ \bibinfo {pages} {488} (\bibinfo {year}
  {2020})}\BibitemShut {NoStop}%
\bibitem [{\citenamefont {Burdette}\ \emph {et~al.}(2020)\citenamefont
  {Burdette} \emph {et~al.}}]{Burdette:2020bke}%
  \BibitemOpen
  \bibfield  {author} {\bibinfo {author} {\bibfnamefont {D.~P.}\ \bibnamefont
  {Burdette}} \emph {et~al.},\ }\href {\doibase 10.1103/PhysRevC.101.055504}
  {\bibfield  {journal} {\bibinfo  {journal} {Phys. Rev. C}\ }\textbf {\bibinfo
  {volume} {101}},\ \bibinfo {pages} {055504} (\bibinfo {year}
  {2020})}\BibitemShut {NoStop}%
\bibitem [{\citenamefont {Long}\ \emph {et~al.}(2020)\citenamefont {Long} \emph
  {et~al.}}]{Long:2020lby}%
  \BibitemOpen
  \bibfield  {author} {\bibinfo {author} {\bibfnamefont {J.}~\bibnamefont
  {Long}} \emph {et~al.},\ }\href {\doibase 10.1103/PhysRevC.101.015501}
  {\bibfield  {journal} {\bibinfo  {journal} {Phys. Rev. C}\ }\textbf {\bibinfo
  {volume} {101}},\ \bibinfo {pages} {015501} (\bibinfo {year}
  {2020})}\BibitemShut {NoStop}%
\bibitem [{\citenamefont {M\"uller}\ \emph {et~al.}(2022)\citenamefont
  {M\"uller} \emph {et~al.}}]{Muller:2022jew}%
  \BibitemOpen
  \bibfield  {author} {\bibinfo {author} {\bibfnamefont {P.}~\bibnamefont
  {M\"uller}} \emph {et~al.},\ }\href {\doibase 10.1103/PhysRevLett.129.182502}
  {\bibfield  {journal} {\bibinfo  {journal} {Phys. Rev. Lett.}\ }\textbf
  {\bibinfo {volume} {129}},\ \bibinfo {pages} {182502} (\bibinfo {year}
  {2022})},\ \Eprint {http://arxiv.org/abs/2206.00742} {arXiv:2206.00742
  [nucl-ex]} \BibitemShut {NoStop}%
\bibitem [{\citenamefont {Long}\ \emph {et~al.}(2022)\citenamefont {Long} \emph
  {et~al.}}]{Long:2022yea}%
  \BibitemOpen
  \bibfield  {author} {\bibinfo {author} {\bibfnamefont {J.}~\bibnamefont
  {Long}} \emph {et~al.},\ }\href {\doibase 10.1103/PhysRevC.106.045501}
  {\bibfield  {journal} {\bibinfo  {journal} {Phys. Rev. C}\ }\textbf {\bibinfo
  {volume} {106}},\ \bibinfo {pages} {045501} (\bibinfo {year}
  {2022})}\BibitemShut {NoStop}%
\bibitem [{\citenamefont {Plestid}(2024)}]{eikonal_algebra}%
  \BibitemOpen
  \bibfield  {author} {\bibinfo {author} {\bibfnamefont {R.}~\bibnamefont
  {Plestid}},\ }\href@noop {} {\  (\bibinfo {year} {2024})},\ \Eprint
  {http://arxiv.org/abs/2402.14769} {arXiv:2402.14769 [hep-ph]} \BibitemShut
  {NoStop}%
\bibitem [{Note18()}]{Note18}%
  \BibitemOpen
  \bibinfo {note} {See Supplemental Material [URL-HERE] for a discussion of how
  these numerical shifts are computed.}\BibitemShut {Stop}%
\bibitem [{\citenamefont {Gennari}\ \emph {et~al.}(2023)\citenamefont {Gennari}
  \emph {et~al.}}]{Gennari:INT}%
  \BibitemOpen
  \bibfield  {author} {\bibinfo {author} {\bibfnamefont {M.}~\bibnamefont
  {Gennari}} \emph {et~al.},\ }\href
  {https://www.int.washington.edu/sites/default/files/schedule_session_files/Gennari_M2.pdf}
  {\enquote {\bibinfo {title} {Standard model corrections to fermi transitions
  in light nuclei},}\ } (\bibinfo {year} {2023}),\ \bibinfo {note} {~INT
  PROGRAM 23-1B, {\it New physics searches at the precision frontier
  }}\BibitemShut {NoStop}%
\bibitem [{Note19()}]{Note19}%
  \BibitemOpen
  \bibinfo {note} {Recall that we are counting $v\cdot v^\prime = E/m$ as order
  unity. Since the electron is pointlike there is no further restriction on
  $v\cdot v^\prime $, unlike the case for the decay of heavy mesons with
  internal structure~\cite {Isgur:1990yhj}.}\BibitemShut {Stop}%
\bibitem [{Note20()}]{Note20}%
  \BibitemOpen
  \bibinfo {note} {Remaining radiative corrections, denoted by $(1+\Delta
  _R^V)(1+\delta _{\protect \rm NS} - \delta _C)$ in Ref.~\cite
  {Hardy:2020qwl}, are identified (up to a prefactor involving $G_F V_{ud}$)
  with the matching coefficient ${\protect \cal C}(\mu _H)$ in the effective
  Lagrangian (\ref {eq:LH}).}\BibitemShut {Stop}%
\bibitem [{\citenamefont {Hardy}\ and\ \citenamefont
  {Towner}(2005)}]{Hardy:2004id}%
  \BibitemOpen
  \bibfield  {author} {\bibinfo {author} {\bibfnamefont {J.~C.}\ \bibnamefont
  {Hardy}}\ and\ \bibinfo {author} {\bibfnamefont {I.~S.}\ \bibnamefont
  {Towner}},\ }\href {\doibase 10.1103/PhysRevC.71.055501} {\bibfield
  {journal} {\bibinfo  {journal} {Phys. Rev. C}\ }\textbf {\bibinfo {volume}
  {71}},\ \bibinfo {pages} {055501} (\bibinfo {year} {2005})},\ \Eprint
  {http://arxiv.org/abs/nucl-th/0412056} {arXiv:nucl-th/0412056} \BibitemShut
  {NoStop}%
\bibitem [{Note21()}]{Note21}%
  \BibitemOpen
  \bibinfo {note} {The Table lists transitions for which the total ${\protect
  \cal F}t$ uncertainty is $\lesssim 10^{-3}$, {\protect \it cf}. Fig.~4 of
  Ref.~\cite {Hardy:2020qwl}.}\BibitemShut {Stop}%
\bibitem [{\citenamefont {Isgur}\ and\ \citenamefont
  {Wise}(1990)}]{Isgur:1990yhj}%
  \BibitemOpen
  \bibfield  {author} {\bibinfo {author} {\bibfnamefont {N.}~\bibnamefont
  {Isgur}}\ and\ \bibinfo {author} {\bibfnamefont {M.~B.}\ \bibnamefont
  {Wise}},\ }\href {\doibase 10.1016/0370-2693(90)91219-2} {\bibfield
  {journal} {\bibinfo  {journal} {Phys. Lett. B}\ }\textbf {\bibinfo {volume}
  {237}},\ \bibinfo {pages} {527} (\bibinfo {year} {1990})}\BibitemShut
  {NoStop}%
\bibitem [{\citenamefont {Beg}\ \emph {et~al.}(1969)\citenamefont {Beg},
  \citenamefont {Bernstein},\ and\ \citenamefont {Sirlin}}]{Beg:1969zu}%
  \BibitemOpen
  \bibfield  {author} {\bibinfo {author} {\bibfnamefont {M.~A.~B.}\
  \bibnamefont {Beg}}, \bibinfo {author} {\bibfnamefont {J.}~\bibnamefont
  {Bernstein}}, \ and\ \bibinfo {author} {\bibfnamefont {A.}~\bibnamefont
  {Sirlin}},\ }\href {\doibase 10.1103/PhysRevLett.23.270} {\bibfield
  {journal} {\bibinfo  {journal} {Phys. Rev. Lett.}\ }\textbf {\bibinfo
  {volume} {23}},\ \bibinfo {pages} {270} (\bibinfo {year} {1969})}\BibitemShut
  {NoStop}%
\bibitem [{\citenamefont {Ji}\ and\ \citenamefont {Musolf}(1991)}]{Ji:1991pr}%
  \BibitemOpen
  \bibfield  {author} {\bibinfo {author} {\bibfnamefont {X.-D.}\ \bibnamefont
  {Ji}}\ and\ \bibinfo {author} {\bibfnamefont {M.~J.}\ \bibnamefont
  {Musolf}},\ }\href {\doibase 10.1016/0370-2693(91)91916-J} {\bibfield
  {journal} {\bibinfo  {journal} {Phys. Lett. B}\ }\textbf {\bibinfo {volume}
  {257}},\ \bibinfo {pages} {409} (\bibinfo {year} {1991})}\BibitemShut
  {NoStop}%
\bibitem [{\citenamefont {Chetyrkin}\ and\ \citenamefont
  {Grozin}(2003)}]{Chetyrkin:2003vi}%
  \BibitemOpen
  \bibfield  {author} {\bibinfo {author} {\bibfnamefont {K.~G.}\ \bibnamefont
  {Chetyrkin}}\ and\ \bibinfo {author} {\bibfnamefont {A.~G.}\ \bibnamefont
  {Grozin}},\ }\href {\doibase 10.1016/S0550-3213(03)00490-5} {\bibfield
  {journal} {\bibinfo  {journal} {Nucl. Phys. B}\ }\textbf {\bibinfo {volume}
  {666}},\ \bibinfo {pages} {289} (\bibinfo {year} {2003})},\ \Eprint
  {http://arxiv.org/abs/hep-ph/0303113} {arXiv:hep-ph/0303113} \BibitemShut
  {NoStop}%
\bibitem [{\citenamefont {Baikov}\ \emph {et~al.}(2012)\citenamefont {Baikov},
  \citenamefont {Chetyrkin}, \citenamefont {Kuhn},\ and\ \citenamefont
  {Rittinger}}]{Baikov:2012zm}%
  \BibitemOpen
  \bibfield  {author} {\bibinfo {author} {\bibfnamefont {P.~A.}\ \bibnamefont
  {Baikov}}, \bibinfo {author} {\bibfnamefont {K.~G.}\ \bibnamefont
  {Chetyrkin}}, \bibinfo {author} {\bibfnamefont {J.~H.}\ \bibnamefont {Kuhn}},
  \ and\ \bibinfo {author} {\bibfnamefont {J.}~\bibnamefont {Rittinger}},\
  }\href {\doibase 10.1007/JHEP07(2012)017} {\bibfield  {journal} {\bibinfo
  {journal} {JHEP}\ }\textbf {\bibinfo {volume} {07}},\ \bibinfo {pages} {017}
  (\bibinfo {year} {2012})},\ \Eprint {http://arxiv.org/abs/1206.1284}
  {arXiv:1206.1284 [hep-ph]} \BibitemShut {NoStop}%
\bibitem [{\citenamefont {Herzog}\ \emph {et~al.}(2017)\citenamefont {Herzog}
  \emph {et~al.}}]{Herzog:2017ohr}%
  \BibitemOpen
  \bibfield  {author} {\bibinfo {author} {\bibfnamefont {F.}~\bibnamefont
  {Herzog}} \emph {et~al.},\ }\href {\doibase 10.1007/JHEP02(2017)090}
  {\bibfield  {journal} {\bibinfo  {journal} {JHEP}\ }\textbf {\bibinfo
  {volume} {02}},\ \bibinfo {pages} {090} (\bibinfo {year} {2017})},\ \Eprint
  {http://arxiv.org/abs/1701.01404} {arXiv:1701.01404 [hep-ph]} \BibitemShut
  {NoStop}%
\bibitem [{\citenamefont {Hill}(2017)}]{Hill:2016gdf}%
  \BibitemOpen
  \bibfield  {author} {\bibinfo {author} {\bibfnamefont {R.~J.}\ \bibnamefont
  {Hill}},\ }\href {\doibase 10.1103/PhysRevD.95.013001} {\bibfield  {journal}
  {\bibinfo  {journal} {Phys. Rev. D}\ }\textbf {\bibinfo {volume} {95}},\
  \bibinfo {pages} {013001} (\bibinfo {year} {2017})},\ \Eprint
  {http://arxiv.org/abs/1605.02613} {arXiv:1605.02613 [hep-ph]} \BibitemShut
  {NoStop}%
\bibitem [{\citenamefont {Czarnecki}\ \emph {et~al.}(2004)\citenamefont
  {Czarnecki}, \citenamefont {Marciano},\ and\ \citenamefont
  {Sirlin}}]{Czarnecki:2004cw}%
  \BibitemOpen
  \bibfield  {author} {\bibinfo {author} {\bibfnamefont {A.}~\bibnamefont
  {Czarnecki}}, \bibinfo {author} {\bibfnamefont {W.~J.}\ \bibnamefont
  {Marciano}}, \ and\ \bibinfo {author} {\bibfnamefont {A.}~\bibnamefont
  {Sirlin}},\ }\href {\doibase 10.1103/PhysRevD.70.093006} {\bibfield
  {journal} {\bibinfo  {journal} {Phys. Rev.}\ }\textbf {\bibinfo {volume}
  {D70}},\ \bibinfo {pages} {093006} (\bibinfo {year} {2004})},\ \Eprint
  {http://arxiv.org/abs/hep-ph/0406324} {arXiv:hep-ph/0406324 [hep-ph]}
  \BibitemShut {NoStop}%
\end{thebibliography}%

\end{document}